\documentclass[preprint,12pt]{elsarticle}

\usepackage{amssymb}
\usepackage{amsmath}

\usepackage{array}
\usepackage{soul}
\usepackage{xcolor}
\usepackage{multirow}
\usepackage[hyphens]{url}
\usepackage{tabularray}
\UseTblrLibrary{booktabs}

\usepackage{pdflscape}

\journal{Computer Networks}

\begin{document}

\begin{frontmatter}



\title{From Simulation to Deep Learning: Survey on Network Performance Modeling Approaches}


\author[]{Carlos Güemes-Palau}
\ead{carlos.guemes@upc.edu}
\author[]{Miquel Ferriol-Galmés, Jordi Paillisse-Vilanova, Pere Barlet-Ros, Albert Cabellos-Aparicio} 

\affiliation{organization={Universitat Politècnica de Catalunya (UPC)},
            city={Barcelona},
            state={Catalonia},
            country={Spain}}

\begin{abstract}
Network performance modeling is a field that predates early computer networks and the beginning of the Internet. It aims to predict the traffic performance of packet flows in a given network. Its applications range from network planning and troubleshooting to feeding information to network controllers for configuration optimization.
Traditional network performance modeling has relied heavily on Discrete Event Simulation (DES) and analytical methods grounded in mathematical theories such as Queuing Theory and Network Calculus. However, as of late, we have observed a paradigm shift, with attempts to obtain efficient Parallel DES, the surge of Machine Learning models, and their integration with other methodologies in hybrid approaches. This has resulted in a great variety of modeling approaches, each with its strengths and often tailored to specific scenarios or requirements.
In this paper, we comprehensively survey the relevant network performance modeling approaches for wired networks over the last decades. With this understanding, we also define a taxonomy of approaches, summarizing our understanding of the state-of-the-art and how both technology and the concerns of the research community evolve over time.
Finally, we also consider how these models are evaluated, how their different nature results in different evaluation requirements and goals, and how this may complicate their comparison.
\end{abstract}




\end{frontmatter}

\tableofcontents


\section{Introduction}
\label{sec:introduction}


Performance modeling has long been a fundamental tool in computer networking. By enabling the prediction and evaluation of network behavior without requiring disruptive changes in live systems, models support network design, planning, and protocol development. From early backbone networks to modern cloud infrastructures, performance models have helped operators make informed, data-driven decisions about scalability, reliability, and efficiency.

In the research community, much of the focus over the past two decades has been on wireless networks, driven by the rapid rise of mobile communication, 5G/6G technologies, and the Internet of Things. Wired networks, by contrast, seemed comparatively static: their role was often limited to large national backbones such as GBN~\cite{5970940} or Abilene~\cite{abeliene}, and for a time, modeling research in this area stagnated.

However, this perception dramatically changed with the rise of data centers. Once motivated mainly by cloud computing and large-scale web services, today’s data centers are increasingly driven by artificial intelligence (AI) and large language model (LLM) training workloads. According to the International Energy Agency~\cite{IEADatacenter}, conventional servers increased their energy demand from 145 TWh in 2020 to 195 TWh in 2024, with projections exceeding 300 TWh by 2030. Accelerated servers designed for AI workloads project even steeper growth, increasing from 10 TWh in 2020 to 60 TWh in 2024, and are similarly expected to surpass 300 TWh by 2030. 
These figures illustrate the unprecedented demands on wired networks, particularly data center networks, and the reason why modeling them has become an urgent research priority.


Unlike domains such as electromagnetism or fluid dynamics, where governing equations capture system behavior, computer networks lack a single set of exact mathematical laws. Their discrete nature, combined with internet traffic's bursty and self-correlated behavior~\cite{Popoola2017EmpiricalModel, Alasmar2021InternetPrediction} and the complexity of congestion control makes accurate modeling a continuous challenge. As networks evolve, so do modeling approaches, from formula-based analytical models and packet-level discrete event simulation to, more recently, machine learning–driven prediction.

From the 1990s to the 2010s, most network performance models were based on analytical modeling or simulation. Analytical models, such as those built on Queuing Theory (QT)~\cite{Jackson1963Jobshop-likeSystems}, described networks as systems of equations. While computationally efficient, they often rely on simplified assumptions about traffic and service distributions. Network Calculus (NC)~\cite{Cruz1991AIsolation, Cruz1991AAnalysis}, alternatively, offered deterministic worst-case performance bounds without assuming specific traffic or service distributions. However, its bounds were often overly conservative and limited to feedforward topologies.
Conversely, Discrete Event Simulation (DES) became popular by offering high-fidelity, packet-level modeling of networks. Yet, its computational cost, proportional to the number of network events, made it impractical for large-scale and high-speed networks. While each approach had its strengths, neither fully met the growing demands for scalability, expressiveness, and accuracy.

\begin{figure}[t]
    \centering
    \includegraphics[width=0.8\columnwidth]{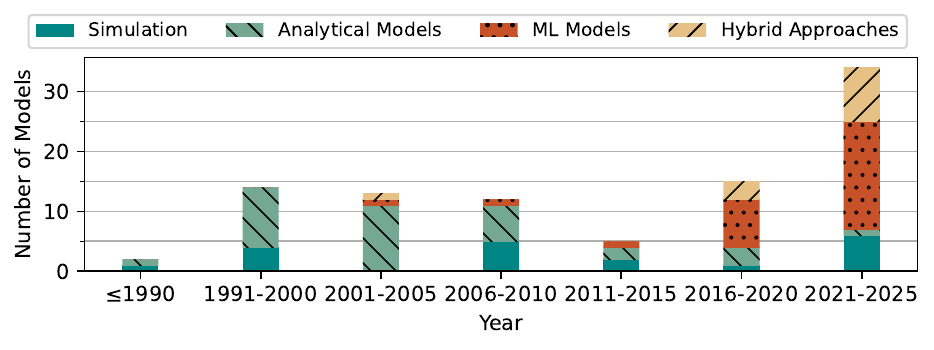}
    \caption{Identified models per year of publication and type. Does not consider other types of publications, such as surveys. If the same model is covered in multiple papers, it is counted once in the year of its earliest publication.}
    \label{fig:evolution}
\end{figure}

After a period of stagnation in the 2010s, research in network performance modeling was revitalized by the rise of Machine Learning (ML). Although the first applications of ML to network modeling date back to 2001, it was the surge of Deep Learning (DL) models in 2018 that brought a significant breakthrough. These models proved capable of accurately predicting network performance while being relatively inexpensive to train and run.
This marked a major shift away from traditional analytical models. Although analytical approaches provide theoretical guarantees, DL models have increasingly outperformed them in practice, providing higher accuracy at similar or even lower computational costs. As a result, DL-based models quickly became the dominant approach in the literature.
Beyond replacing earlier techniques, DL also inspired a second paradigm shift: the emergence of hybrid approaches. Researchers began integrating ML to existing approaches to leverage their complementary strengths. For instance, ML models have been incorporated into DES frameworks to reduce simulation time without sacrificing detail or expressiveness. 

This trend is reflected in Figure~\ref{fig:evolution}, which shows the number of network performance models published over the past decades. Research activity peaked in the late 1990s and early 2000s, followed by a period of stagnation around 2010. However, since 2018, the field has experienced a strong resurgence, with ML-based models becoming the dominant approach. The figure also shows the steady rise of hybrid methods and a decline in the use of traditional analytical models. 
In contrast, the number of simulation-based models has remained relatively stable. Although simulation remains a reliable and expressive modeling tool, recent research has shifted toward improving its scalability and execution speed. 

In this survey, we study the evolution of network performance models over time. We analyze the motivations behind each shift, the limitations that shaped subsequent approaches, and the emerging trends in the latest generation of models. Additionally, we highlight the challenges the field currently faces and the directions research is taking to address them.
While surveys exist in related areas, such as wireless networks~\cite{8382166, 10198239, Verdecchia2024NetworkReview} or specific modeling approaches like NC~\cite{5415864, 10.5555/1434872} and Graph Neural Networks~\cite{JIANG202240}, to the best of our knowledge, there has not been a dedicated survey that revisits performance modeling with a focus on wired networks.

\subsection{Methodology and scope}
To find all relevant models published in the state-of-the-art (from now on referred to as SotA), we searched within IEEE Xplore, ACM's Digital Library, and Elvisier's ScienceDirect for papers with either the terms ``network modeling" or ``network performance". We focused on papers published within the IEEE INFOCOM, IEEE NOMS, IEEE/ACM Transactions on Networking, ACM SIGCOMM, ACM CoNEXT, Elsevier Computer Networks, and Elsevier Computer Communications conferences and journals. We then excluded papers that fell outside the scope of the survey and papers that lacked a network model capable of predicting performance metrics.
For certain high-impact papers, such as \cite{Zhang2021MimicNet:Learning, Yang2022DeepQueueNet:Visibility, Ferriol-Galmes2023RouteNet-Fermi:Networks, Gao2023DONS:Parallelization}, we also reviewed articles that cite them. For older methods, such as the original QT models, we started by searching in the references of the papers we had already found, and then repeated this process recursively until finding their original precursor. 

This survey limits itself to models capable of replicating or predicting the performance of traffic flows within wired networks. While there are similar research fields, ultimately, the solutions they encompass attempt to solve differently-natured problems, rendering their comparison fruitless. This applies to the following fields:
\begin{itemize}
    \item \textbf{Wireless and mobile networks}: The methodologies for modeling wired and wireless networks vary significantly due to their differences. Hence, in this survey, we focus only on wired networks, and we refer the interested reader to existing surveys on wireless networks \cite{8382166, 10198239, Verdecchia2024NetworkReview}.
    \item \textbf{Network verification}: Rather than predicting network behavior under specific conditions, these models focus on ensuring the viability of potential network configurations (e.g., if certain QoS guarantees are maintained~\cite{Arun2021TowardBehavior, Arashloo2023FormalAnalysis}).
    \item \textbf{Traffic prediction models}: Unlike network modeling, whose task is to predict the performance of traffic flows within the network, in traffic prediction, the task is to predict and describe the incoming traffic flows.
    \item \textbf{Anomaly detection}: These models are tasked with detecting anomalous patterns that can be indicative of a network security threat. Rather than making predictions, they discriminate between the ``normal" expected behaviors and the potentially dangerous unexpected ones.  
\end{itemize}

\subsection{Describing Network Models}

Throughout the survey, we will summarize the identified models in tables. In the following section, we describe how we identify models:

\subsubsection*{Model Type} Specific type of model (e.g., if simulation, type of simulator; if ML model, which ML architecture...). 

\subsubsection*{Input Scope}
Refers to the scope of the input data of the model:
\begin{itemize}
    \item Single-flow scenarios: models can only model single flows. They do not consider cross-traffic interactions. They also ignore the underlying network topology, or only consider individual links or devices.
    \item Traffic-matrix scenarios: models consider cross-traffic interactions. All flows are explicitly represented, even if the model produces predictions for just a subset of them. However, they still heavily simplify or ignore the underlying network topology.
    \item Full network scenarios: models consider cross-traffic interactions, and also consider the network's topology, routing, and characteristics in their calculations.
\end{itemize}

\subsubsection*{Output Scope}
Refers to the scope of the predicted performance metrics:
\begin{itemize}
    \item Flow-level: average performance metrics for each flow.
    \item Flow-level with temporal component: like before, but with a temporal component. Methods may vary in the coarseness of the temporal component or how time is aggregated.
    \item Packet-level: can predict individual packets.
\end{itemize}
    
\subsubsection*{Supported Traffic Types}
\begin{itemize}
    \item UDP: Traffic without congestion control.
    \item TCP: Traffic regulated by congestion control. Some models may consider a generic congestion control algorithm; others consider one or multiple versions of TCP.
    \item Non-specific: The model does not specify which traffic distributions it supports. However, because of the nature of how they model traffic distributions, this may result in varying degrees of error depending on the packet arrival time distribution (e.g., queue models modeling traffic through a Poisson process)
    \item Any: can support faithfully any traffic type. 
\end{itemize}

\subsubsection*{Supported Performance Metrics}
These usually depend on the output scope and traffic type supported by the network model:
\begin{itemize}
    \item Models with packet-level predictions usually predict the full packet information (i.e., the timestamp of the packet at each point of its routing path)
    \item Models for UDP traffic focus on packet delay, jitter, and loss rates.
    \item Models for TCP traffic may also focus on flow completion time (FCT), throughput, and packet round-trip time (RTT).
\end{itemize}

\subsubsection*{Evaluation}
Most network models are evaluated or offer theoretical guarantees. If they are, we categorize evaluation in the following manner:
\begin{itemize}
    \item Analytical evaluation: properties are analyzed from an analytical viewpoint only (i.e., verified through formal proofs). Common in analytical models. Restricted to assumptions made by the analysis itself.
    \item Simulated data evaluation: The model was evaluated using simulated data against the simulator's ground truth.
    \item Testbed data evaluation: The model was evaluated using data generated and captured from a testbed network.
    \item Real data evaluation: The model was evaluated using captured real internet data. 
\end{itemize}

\subsection{Survey Outcomes}

The main contributions of this survey are as follows:
\begin{itemize}
    \item We introduce a taxonomy for describing the different kinds of network performance models, allowing us to better understand the SotA and how the various approaches influence and complement each other.
    \item We perform a comprehensive study of the relevant network performance models present in the SotA. 
    \item We discuss the trends and limitations of the current SotA of network performance modeling, comparing them to those identified by researchers back at the start of the millennium.
\end{itemize}

\subsection{Structure of Survey}

\begin{table}[!hp]
\centering
\begin{tblr}{
  width=\textwidth,
  colsep=3pt, rowsep=0.1pt,
  colspec={X[r,0.2] X[l,1]},
  rulesep=0pt,
  stretch=0.95,
  row{1} = {font=\bfseries},
  hline{1} = {1pt},
  hline{2}   = {0.6pt},
}
Acronym         & Description                           \\ \midrule
ADNC         & Algebraic-based Deterministic Network Calculus                           \\
AI           & Artificial Intelligence \\
AQM          & Active Queue Management                                                  \\
CVAE         & Conditional Variational Auto-Encoder                                      \\
{[P]}DES       & {[Parallel]} Discrete Event Simulation                                     \\
DCQCN          & Data Center Quantized Congestion Notification~\cite{10.1145/2829988.2787484}                                                            \\
DL           & Deep Learning                                                            \\
DNN          & Dense Neural Network                                                     \\
EVT          & Extreme Value Theory                                                     \\
FCT          & Flow Completion Time                                                     \\
FIFO         & First-In First-Out                                                       \\
GCN          & Graph Convolutional Network~\cite{kipf2016semi}                          \\
GGNN         & Gated Graph Neural Network~\cite{li2016gated}                            \\
GNN          & Graph Neural Network~\cite{4700287}                                      \\
IPG          & Inter-Packet Gap                                                         \\
LLM          & Large Language Model \\
LP           & Logical Process                                                          \\
LPP          & Linear Programming Problem                                                \\
{[Bi]}LSTM      & {[Bidirectional]} Long Short-Term Memory~\cite{10.1162/neco.1997.9.8.1735} \\
{[EW]}MA       & {[Exponentially Weighted]} Moving Average \\
ML           & Machine Learning                                                         \\
MPNN         & Message-Passing Neural Network~\cite{pmlrv70gilmer17a}                   \\
NC           & Network Calculus                                                         \\
NP           & Neural Processes~\cite{garnelo2018neuralprocesses}                       \\
ns{[-2$\|$-3]} & The Network Simulator~\cite{1995TheNs-2,Riley2010TheSimulator}           \\
ODE          & (System of) Ordinary Differential Equations                              \\
ODNC         & Optimization-based Deterministic Network Calculus                        \\
PBOO         & Pay Bursts Only Once~\cite{2001NetworkCalculus}          \\
PDE          & (System of) Partial Differential Equations                               \\
PMOO         & Pay Multiplexing Only Once~\cite{Fidler2003ExtendingScheduling}          \\
QoS          & Quality of Service                                                       \\
QT           & Queuing Theory                                                           \\
RBFNN        & Radial Basis Function Neural Network \\
RED          & Random Early Detection                                                   \\
RF           & Random Forest                                                            \\
RGCN          & Relational Graph Convolutional Network~\cite{schlichtkrull2018modeling} \\
RNN & Recurrent Neural Network \\
RON & Resilient Overlay Networks~\cite{RONtestbed} \\
\end{tblr}%
\end{table}

\begin{table}[t!]
\centering
\begin{tblr}{
  width=\textwidth,
  colsep=3pt, rowsep=0.1pt,
  colspec={X[r,0.2] X[l,1]},
  rulesep=0pt,
  stretch=0.95,
}
RON & Resilient Overlay Networks~\cite{RONtestbed} \\
RTT          & Round Trip Time                                                          \\
SDE          & (System of) Stochastic Differential Equations                            \\
SFA          & Separate Flow Analysis~\cite{2001NetworkCalculus}                        \\
SNC          & Stochastic Network Calculus                                              \\
SVR          & Support Vector Regression                                                \\
{[DC]}TCP          & {[Data Center \cite{10.1145/1851182.1851192}]} Transmission Control Protocol                                            \\
TFA          & Total Flow Analysis~\cite{Cruz1991AIsolation, Cruz1991AAnalysis}         \\
TMA          & Tandem Matching Analysis~\cite{Bondorf2017QualityAnalysis}               \\
UDP          & User Datagram Protocol                                                   \\
\hline
\end{tblr}%
\caption{List of acronyms}
\label{tab:acronyms}
\end{table}

For ease of reading, we summarize the structure of the survey. We also include a summary of the acronyms used throughout the survey in Table~\ref{tab:acronyms}.
\begin{itemize}
    \item In Section~\ref{sec:taxonomy} we introduce and describe our taxonomy for network performance models.
    \item In the main body of the paper, we survey the current or previously relevant network performance models. Specifically, Section~\ref{sec:simulation} covers network simulators, Section~\ref{sec:analytical_models} analytical models, Section~\ref{sec:ml_models} ML models, and Section \ref{sec:hybrid_models} hybrid approaches.
    \item Section~\ref{sec:discussion} includes our discussion of the SotA of network performance models.
    \item Finally, in Section~\ref{sec:future} we discuss future directions of research into network performance models.
\end{itemize}

\section{Taxonomy of Network Performance Models}
\label{sec:taxonomy}
\begin{figure*}[t]
    \centering
    \includegraphics[width=\textwidth]{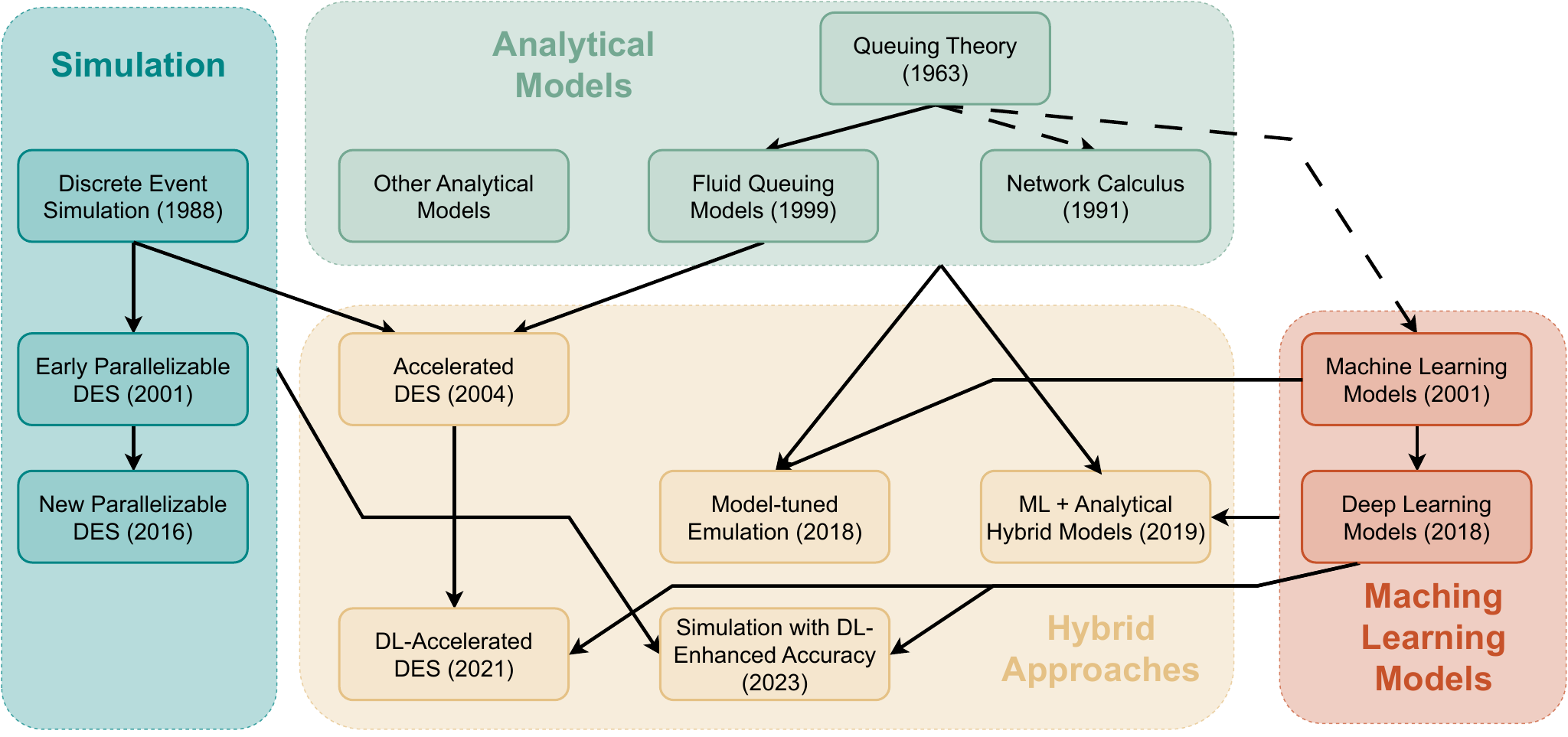}
    \caption{Taxonomy of network performance models. Solid line arrows indicate direct evolution, while dashed line arrows indicate inspiration or a ``response to". Year indicates the year of the earliest publication within that category.}
    \label{fig:taxonomy}
\end{figure*}

The evolution of network performance models is summarized in Figure~\ref{fig:taxonomy}. This taxonomy shows the four main types of network models, simulation, analytical models, ML models, and hybrid approaches, and how they interact.

The earliest network performance models were analytical, based on Queuing Theory (QT)~\cite{Jackson1963Jobshop-likeSystems}. These models describe the behavior of packets in devices through systems of queues, based on given assumptions (e.g., packet arrival distributions). Simpler QT models, constrained by stricter assumptions, are often inaccurate. More advanced models relax these assumptions and improve accuracy, but become significantly harder to solve. Additionally, QT models typically provide only aggregate predictions, such as average packet delay, rather than detailed, per-packet insights. This makes them better suited for general approximations of network performance, rather than thoroughly examining specific, complex scenarios.

This has inspired other types of analytical models, intending to improve accuracy, cost, and expressiveness compared to their QT counterparts. On the one hand, fluid queuing models iterate over traditional discrete QT models while assuming that transmitted data can be represented as a continuous value. While this assumption does not hold in reality, it allows for network performance models built through systems of differential equations, which can be solved efficiently. Hence, these models are inexpensive, while the inclusion of the temporal component in the differential equations also makes them more expressive.

Alternatively, other researchers turn to Network Calculus (NC)~\cite{Cruz1991AIsolation, Cruz1991AAnalysis}. Unlike QT, initial NC models aimed to reduce the number of assumptions, include the temporal component, and provide predictable and theoretically proven correct performance predictions at a low computational cost. This is achieved by offering worst-case bounds for the packet delay as a function of time. However, while the bounds are proved to be correct, early NC models also show that these tend to be too pessimistic, while more complex NC models prove to be too computationally expensive to be practical.

By contrast, an old yet popular alternative for building network models is simulation, specifically Discrete Event Simulation (DES). It allows for accurate, fine-grained results. Popular DES projects like OMNET~\cite{Varga2010OMNeT++} and ns~\cite{1995TheNs-2} are open source, allowing researchers to share the implementations for newer protocols and devices and facilitating their use. The most significant limitation of DES is its computational cost, which is compounded by the fact that its sequential nature makes it extremely hard to parallelize.
As a result, the development of parallelizable DES (PDES) became a prominent research area, driven both by the technical challenges involved and the potential benefits: an accurate, detailed modeling approach that could scale to large networks if implemented effectively.

QT and other analytical models also influenced the introduction of the first ML models. Originally, the researchers referred to the former as ``formula-based" and the latter as ``history-based" \cite{He2005OnThroughput}. The objective was for these models to offer better accuracy-cost trade-offs compared to the analytical models. However, ML models would not start to thrive until 2018, with the introduction of DL. Newer architectures, such as Graph Neural Networks~\cite{4700287}, were expressive enough to extract and process as much information as was available and produce more accurate results. Since their introduction, DL models have dominated, completely overtaking analytical models as they offered more accurate performance predictions at similar or lower computational costs. Furthermore, beyond building ML or DL models directly, there is also an interest in integrating them into previously researched approaches.

We refer to these as hybrid approaches and are, by their nature, extremely varied as to how they may achieve this. For example, ML and analytical models can be combined by using the output predictions of one approach as input to the other and obtain an overall more accurate prediction than if done separately \cite{deAquinoAfonso2022QT-Routenet:Theory, Helm2023PredictingGNNs}. Another approach is to augment network emulators, useful for network verification but not for performance prediction, with queuing models and Bayesian optimization to enable accurate performance prediction~\cite{Yan2018Pantheon:Research, Ashok2022Data-DrivenIBox}. One of the most promising directions involves enhancing DES by integrating fluid models and, more recently, DL. This approach aims to support parallelization and lower simulation costs by replacing selected components with alternative models, while preserving the high accuracy and expressiveness of traditional DES \cite{YuGu2004OnSimulation, Zhang2021MimicNet:Learning}.
Finally, recent work has explored using DL models to enhance simulation accuracy in situations where they are unable to predict network behavior faithfully~\cite{Alomar2023CausalSim:Simulation, Spath2024Sim2HW:Measurements}.


\section{Network Simulation}
\label{sec:simulation}

This section focuses on network models that are implemented through simulation. These are summarized in Table~\ref{tab:simulation}.

\subsection{Discrete Event Simulation}

Discrete Event Simulation describes the scenario through a global state and a sequence of ``discrete events" over time~\cite{Guizani2010NetworkSimulation}. These events are defined as points in time at which the state of the simulation (i.e., the network and the traffic within it) changes. It includes the generation of packets being introduced into a link, reaching the other side and being buffered, to then being queued and processed by the device. The simulator models the state of the network devices and the rules to process the different events by studying, and sometimes even re-implementing their protocols. Protocols can be studied through their specification (e.g., RFC publications).

\begin{landscape}

\begin{table*}[!hp] 
\centering
\small
\begin{tblr}{
  width=\columnwidth,
  colsep=3pt, rowsep=2pt,
  colspec={X[l,0.9] X[l,1.0] X[c,0.9] X[c,0.9] X[c,1.2] X[j,1] X[c,1.1]},
  rulesep=0pt,
  stretch=0.95,
  row{1} = {font=\bfseries},
  hline{1,Z} = {1pt},
  hline{2}   = {0.6pt},
}
Model &
  Type &
  Input scope &
  Output scope &
  Traffic type &
  Performance metrics &
  Evaluation \\ \midrule
REAL~\cite{Keshav1988REAL:Simulator} &
  DES &
  Full network scenarios &
  Packet-level &
  Any &
  Full packet path information &
  \\ \midrule
ns~\cite{1995TheNs-2}, -2, -3~\cite{Riley2010TheSimulator} &
  DES &
  Full network scenarios &
  Packet-level &
  Any &
  Full packet path information &
   \\ \midrule
OMNeT, OMNeT++~\cite{Varga2010OMNeT++} &
  DES &
  Full network scenarios &
  Packet-level &
  Any &
  Full packet path information &
   \\ \midrule
OPNET~\cite{Chang1999NetworkOPNET} &
  DES &
  Full network scenarios &
  Packet-level &
  Any &
  Full packet path information &
   \\ \midrule
CaSiNo~\cite{Guizani2010NetworkSimulation} &
  DES &
  Full network scenarios &
  Packet-level &
  Any &
  Full packet path information &
   \\ \midrule
WNS~\cite{BultmannDaniel2010OpenWNS} &
  DES &
  Full network scenarios &
  Packet-level &
  Any &
  Full packet path information &
   \\ \midrule
IKR Simulation Library~\cite{Sommer2010IKRLibrary} &
  DES &
  Full network scenarios &
  Packet-level &
  Any &
  Full packet path information &
   \\ \midrule
Java modeling tools~\cite{Casale2011QuantitativeTools} &
  DES &
  Full network scenarios &
  Packet-level &
  Any &
  Full packet path information &
   \\ \midrule
gem5~\cite{Binkert2011TheSimulator} &
  DES &
  Full network scenarios &
  Packet-level &
  Any &
  Full packet path information &
   \\ \midrule
\midrule
\end{tblr}%
\end{table*}

\begin{table*}[!hp] 
\centering
\small
\begin{tblr}{
  width=\columnwidth,
  colsep=3pt, rowsep=2pt,
  colspec={X[l,0.9] X[l,1.0] X[c,0.9] X[c,0.9] X[c,1.2] X[j,1] X[c,1.1]},
  rulesep=0pt,
  stretch=0.95,
  hline{1,Z} = {1pt},
}
\midrule
pd-gem5~\cite{Alian2016Pd-gem5:Systems} &
  DES &
  Full network scenarios &
  Packet-level &
  Any &
  Full packet path information &
  Testbed data \\ \midrule
SimBricks~\cite{Li2022SimBricks:Simulation} &
  Modular DES &
  Full network scenarios &
  Packet-level &
  Any &
  Full packet path information &
  Simulated data \\ \midrule
SplitSim~\cite{Li2024SplitSim:Research} &
  Modular DES &
  Full network scenarios &
  Packet-level &
  Any &
  Full packet path information &
  Simulated data \\ \midrule
DONS~\cite{Gao2023DONS:Parallelization} &
  DES &
  Full network scenarios &
  Packet-level &
  Any &
  Full packet path information &
  Simulated data \\ \midrule
Unison~\cite{Bai2024Unison:Kernel} &
  DES &
  Full network scenarios &
  Packet-level &
  Any &
  Full packet path information &
  Simulated data \\ \midrule 
NSX~\cite{Khashab2025NSX:Server} &
  DES &
  Full network scenarios &
  Packet-level &
  Any &
  Full packet path information &
  Only evaluates cost, not accuracy \\ \midrule
Parsimon~\cite{Zhao2022ScalableNetworks} &
  Link-level DES &
  Full network scenarios &
  Flow-level &
  TCP (Focuses on DCTCP, but generalizable) &
  Tail FCT (90th percentile or higher) &
  Simulated data \\ \midrule
\end{tblr}%
\caption{Summary of simulated network performance models}
\label{tab:simulation}
\end{table*}

\end{landscape}

The earliest dedicated network DES simulator is REAL~\cite{Keshav1988REAL:Simulator}, originally built to evaluate the performance of queuing algorithms in gateways. It then acted as a basis for the more general purpose ``The Network Simulator" or ns~\cite{1995TheNs-2}. ns has been updated over the years, reaching v2.0 (known as ns-2) in 1997, and later v3 (ns-3) in 2010~\cite{Riley2010TheSimulator}. Along with OMNeT (later to become OMNeT++)~\cite{Varga2010OMNeT++}, they are the most widely used network simulators in academia since their creation in the late 1990s.

Note that, during their development, many alternative DES software were released. This includes OPNET~\cite{Chang1999NetworkOPNET}, CaSiNo~\cite{Guizani2010NetworkSimulation}, Open WNS~\cite{BultmannDaniel2010OpenWNS}, IKR Simulation Library~\cite{Sommer2010IKRLibrary}, Java modeling tools~\cite{Casale2011QuantitativeTools} and gem5~\cite{Binkert2011TheSimulator}.
Ultimately, ns and OMNeT remained the most popular choices due to two factors. First, they are open-source projects with continuous development over the years, allowing them to stay updated and support new technologies. Second, both have permissive licenses that allow them to be used freely for research purposes: ns uses the GNU GPL, and OMNeT uses the Academic Public license.

The main advantage of DES is its completeness: by fully simulating the individual interactions in the network, the simulator can return accurate and complete descriptions of the traffic behavior. It can also be used to understand cross-traffic interactions, the impact of the network topology and routing, and congestion control protocols. Because of its accuracy and level of detail, DES is the preferred choice of network operators whenever applicable. Furthermore, as we explore other alternatives of network models, one quickly realizes how DES is widely used as a baseline (discussed later in Section~\ref{sec:simulation_evaluation}). In addition, in the case of models that undergo some ``training" process, DES is commonly used as a source of the training scenarios due to its ability to generate large quantities of scenarios, specifically those too hard to capture in real-life networks.

However, its completeness also results in its main drawback: computational cost. The cost of DES is proportional to the number of events in the network and, in turn, to the amount of traffic in the network, which becomes unmanageable in larger, high-capacity networks \cite{Guemes-Palau2025RouteNet-Gauss:Learning, Zhang2021MimicNet:Learning, Yang2022DeepQueueNet:Visibility}. Furthermore, this is exacerbated by the fact that DES simulators, due to the sequential nature of the events, are usually implemented as single-thread programs (we cover PDES and its challenges in the next subsection).

Furthermore, sometimes simulation software fails to accurately describe certain protocols or network devices. This may be because the implementation of certain devices is unknown, protected by intellectual protection laws, so their behavior cannot be accurately replicated by simulators~\cite{Guemes-Palau2025RouteNet-Gauss:Learning, Yan2018Pantheon:Research}. Similarly, as we discuss later in Section~\ref{sec:discussion}, the evolution of internet traffic and algorithms used may result in different implementations of protocols such as TCP, and the release of new versions that may yet to be supported by the simulation software \cite{Floyd2001DifficultiesInternet, 10.1145/268437.268737}. These factors result in gaps in the simulation software, even becoming outdated if not continuously maintained. Conversely, to aid future updates, the simulators end up with a modular architecture, allowing new components to interact with existing ones, expanding the simulator's applicability over time.

\subsection{Parallel Discrete Event Simulation}
\label{sec:pdes}

Because of the high computational cost, there has always been an incentive to develop Parallel Discrete Event Simulation (PDES)~\cite{Fujimoto2001ParallelSystems, Kunz2010ParallelSimulation}. This consists of dividing and processing the sequence of events in parallel while maintaining the global state between events. While this does not reduce the computational effort (if anything, the required synchronization mechanisms will require an additional effort), it does reduce simulation time by splitting the load between multiple processes.

Most PDES work by splitting the simulation into Logical Processes (LPs). That is, the simulation itself is split across these LPs, a process done manually, where each one runs as if it were its own simulation. Ideally, the LPs are selected to balance the number of events while minimizing inter-process communication, which is much slower than events handled fully locally.
To maintain the global order of events, a synchronization algorithm is required. While many approaches exist, these can be categorized as either conservative or optimistic \cite{Jafer2013SynchronizationSimulation}. Conservative synchronization algorithms attempt to minimize causality errors, i.e., when the LPs process events out of order, necessitating a rollback. Optimistic synchronization algorithms, in contrast, try to maximize the degree of parallelization at a higher risk. Ultimately, many PDES simulators offer both options, as performance may vary depending on the scenario.

While the benefits of PDES are evident, its complexities have prevented it from completely replacing DES. The inter-process messaging and synchronization costs introduce significant overhead while reducing the effective parallelization. Correctly defining the LPs is cumbersome,  making their effective use more difficult~\cite{Gao2023DONS:Parallelization}.
Furthermore, most common networks are hard to partition to begin with —even if the topology is easily partitioned (e.g., a fat tree topology split into its branches), there will still be large volumes of traffic that cross these boundaries. Overall, networks have proven hard to parallelize, with the gains obtained from parallelization outweighed by the overheads present in PDES. 
Still, many simulators, including ns-3 and OMNeT++, have added support for PDES. There have also been developments of PDES inspired by existing DES, like pd-gem5~\cite{Alian2016Pd-gem5:Systems} based on gem5~\cite{Binkert2011TheSimulator}.

Recently, novel approaches have been developed to further reduce PDEs' overhead. SimBricks~\cite{Li2022SimBricks:Simulation} introduces the concept of modular simulation: basically, it integrates other simulators to model different aspects of the network, such as OMNeT++ and ns-3 of the network itself and gem5 for the hosts, for example. This both exploits the features awarded by each integrated simulation software and increases speed by running each simulator in parallel. However, the authors themselves identify that their simulator cannot accelerate individual components. This leads to the simulation being bottlenecked by the slowest components~\cite{Li2024SplitSim:Research}. Furthermore, requiring interoperability between simulators results in some of their features being lost (e.g., gem5's atomic memory protocol) and limits the scenarios SimBricks can cover, such as only simulating single-core hosts.

Some of these issues are addressed in its expansion,  SplitSim~\cite{Li2024SplitSim:Research}. Extra features include support for mixed fidelity simulation, reducing the accuracy of certain components to lower costs; decomposition of the slowest components; and an improved synchronization algorithm. It also simplifies the user configuration through an orchestration framework.

Another novel solution is the implementation of DONS~\cite{Gao2023DONS:Parallelization}. Unlike other simulators, DONS follows a Data-Oriented design in its implementation, restructuring how memory is organized and accessed to improve parallelization. However, these benefits only apply to thread-based parallelization in multi-core processors, but not to multiple processes across different machines. Hence, to facilitate distributed execution, it utilizes an automatic LP partitioning algorithm. DONS prioritizes correctness, both by mathematically proving the robustness of its Data-Oriented design and by employing a conservative synchronization algorithm.
However, DONS's biggest limitation is its Data-Oriented design, making it incompatible with other PDES implementations. This requires all the simulation logic to be re-implemented to fit the new paradigm~\cite{Bai2024Unison:Kernel}.

Conversely, Unison~\cite{Bai2024Unison:Kernel} follows a more standard approach. Its novelty arises from its automatic fine-grained LP partition, easing configuration and improving the efficacy of the selected LPs, and from dynamic scheduling to avoid bottlenecks. It can also reuse frameworks meant for other DES simulators like ns-3. While effective, it also comes with important limitations: its automatic LP partition does not support stateful links (e.g., wireless channels), and its load balancing assumes that all the processors it runs on are identical.
A similar, recent approach is that of NSX~\cite{Khashab2025NSX:Server}. Unlike other PDES, NSX is meant to run in Graphical Processing Units (GPUs). The primary motivation behind this choice is to capitalize on the current popularity of GPU-heavy data centers designed for LLMs. NSX is designed for GPUs' extreme levels of parallelization: for example, using local event queues to maintain a ``local" event order. The main drawback of this approach is presuming the availability of plentiful, powerful GPU hardware, which is monetarily expensive.

An alternative approach to accelerate DES to those presented earlier is Parsimon~\cite{Zhao2022ScalableNetworks}. Parsimon decreases computational cost while enabling efficient parallelization by decomposing a network topology into its links. That is, for each link in a flow's path, Parsimon studies its experienced delay through a set of independent simulations. Each link simulation predicts link behavior by approximating the expected experienced load in the original topology but in a smaller scenario. These link-level simulations are designed to light, and, as they are independent, they can be executed in parallel. Parsimon also offers the option of using a greedy clustering algorithm to share simulation results between similar links, thus reducing the number of them to be executed, albeit at the cost of accuracy.
However, the link simulations introduce error (usually overestimating delays) and are incompatible with packet-level visibility.
Instead, Parsimon can only predict performance metrics that can be computed as the aggregation of the performance at each link, such as the FCT. Because it tends to overestimate, Parsimon is better suited for tail-prediction —i.e., upper bound approximations.



Overall, building an efficient PDES is non-trivial. Due to the sequential nature of the DES, the authors ultimately need to make some concessions to maximize the efficiency of the simulation's parallelization. Yet, when successfully applied, PDES achieves accurate and complete results with low inference times, assuming the users have the hardware to run it on. 

\subsection{Summary}

Overall, simulation, and in particular DES, have been the most complete network model available. It remains the most accurate option, and continuous support both by the research community and  industry allows DES simulators such as OMNeT++~\cite{Varga2010OMNeT++} and ns-3~\cite{Riley2010TheSimulator} to keep up with the advances in modern networks. Its biggest threat, however, is the computational cost. Traditional, single-threaded DES cannot simulate large networks or modern data centers, and even if it can, it is at a prohibitively high computational cost.

While interest in PDES existed ever since the inception of DES itself, its technical challenges were not addressed properly until fairly recently, with examples like SplitSim~\cite{Li2024SplitSim:Research}, DONS~\cite{Gao2023DONS:Parallelization} and Unison~\cite{Bai2024Unison:Kernel}. Even then, it is important to note that PDES (except for Parsimon~\cite{Zhao2022ScalableNetworks}) does not reduce the computational cost of simulation (if anything, it will increase it due to the synchronization overhead), but rather offloads it to mutiple machines. Hence, without the hardware available to run them, simulating large network scenarios are still out of reach.

\section{Analytical Models}
\label{sec:analytical_models}
This section focuses on analytical network models, summarized in Table~\ref{tab:analytical}. These are generally based on some formal theory and are implemented through systems of equations.

\subsection{Queuing Theory}

Queuing Theory (QT) is a field of mathematics that studies the behavior of queue-like systems. In them, a system will receive requests and service them according to a specified distribution (e.g., Poisson). If the system is occupied, requests will be queued until they can be processed. The original model was proposed in the 1960s in \cite{Jackson1963Jobshop-likeSystems}, and was later refined for computer networks in \cite{Kobayashi1977QueueingAnalysis}.
With them, devices and protocols can be approximated through these models to extract relevant performance metrics such as the mean queuing delay.

Queues' definitions vary in complexity~\cite{Guizani2010NetworkSimulation}. Simpler queuing models can be solved instantly, but are limited to simple behaviors and therefore have more stringent assumptions. For example, the simplest queue is the M/M/1 queue, which assumes (1) a Poisson arrival process of packets, (2) exponentially distributed service times, and (3) a single First-In, First-Out (FIFO) queue with infinite buffer. These properties make calculating factors like average package queuing delay or service time trivial, but are also unrealistic. While the infinite buffer is the clearest example, an important assumption is that of a Poisson arrival process. In practice, internet traffic distributions have been shown to be self-correlated and fit best heavy-tail distributions like Weibull \cite{Popoola2017EmpiricalModel} and log-normal \cite{Alasmar2021InternetPrediction} distributions.

\begin{landscape}

\begin{table*}[!hp] 
\centering
\small
\begin{tblr}{
  width=\columnwidth,
  colsep=3pt, rowsep=2pt,
  colspec={X[l,0.9] X[l,1.0] X[c,0.9] X[c,0.9] X[c,1.2] X[j,1.3] X[c,0.8]},
  rulesep=0pt,
  stretch=0.95,
  row{1} = {font=\bfseries},
  hline{1,Z} = {1pt},
  hline{2}   = {0.6pt},
}
Model &
  Type &
  Input scope &
  Output scope &
  Traffic type &
  Performance metrics &
  Evaluation \\ \midrule 
\cite{Kobayashi1977QueueingAnalysis} &
  M/M/1, M/G/$\infty$, M/G/1 queues &
  Single-flow scenarios &
  Flow-level &
  Non-specific &
  Average packet delay, buffer overflow probability &
  Analytical evaluation \\ \midrule 
\cite{Poon2001} &
  M/G/$\infty$ queue &
  Traffic-matrix scenarios &
  Flow-level &
  Non-specific &
  Loss rate &
  Simulated data evaluation \\ \midrule 
\cite{Fiorini2025} &
  GI/G/1 queue &
  Single-flow scenarios &
  Flow-level &
  Non-specific &
  Upper and lower bounds, average packet delay &
  Simulated data evaluation \\ \midrule 
\cite{Garetto2001AFlows} &
  System of M/M/$\infty$ queues &
  Traffic-matrix scenarios &
  Flow-level &
  TCP (Tahoe) &
  Average packet [loss probability, RTT] &
  Simulated data evaluation \\ \midrule 
\cite{Garetto2004ClosedFlows} &
  System of M/G/$\infty$ queues &
  Traffic-matrix scenarios &
  Flow-level &
  TCP (Tahoe) &
  Average packet [loss probability, RTT] &
  Simulated data evaluation \\ \midrule 
\cite{Yu2010ATraffic} &
  Modified M/M/1 queues &
  Traffic-matrix scenarios &
  Flow-level &
  Non-specific &
  Flow throughput, average packet delay &
  Simulated data evaluation \\ \midrule 
\cite{Bonald1999ComparisonFairness} &
  ODE fluid queuing model &
  Single-flow scenarios &
  Flow-level (temporal) &
  TCP (Reno and Vegas) &
  Flow throughput &
  No baseline $^{\mathrm{1}}$ \\ \midrule 
AIMD \cite{Baccelli2002AIMDTraffic} &
  ODE fluid queuing model &
  Traffic-matrix scenarios &
  Flow-level (temporal) &
  TCP (Generic) &
  Flow throughput and mean RTT per sesion &
  Simulated data evaluation \\ \midrule 
\cite{Bohacek2003ATransmission} &
  ODE fluid queuing model &
  Traffic-matrix scenarios &
  Flow-level (temporal) &
  TCP (Generic) &
  Average RRT and loss rate &
  Simulated data evaluation \\ \midrule 
\end{tblr}
\begin{tabular}{m{\columnwidth}}
    $^{\mathrm{1}}$ Analytical comparison between TCP versions, no comparison with baseline. \\
\end{tabular}
\end{table*}

\begin{table*}[!h] 
\centering
\small
\begin{tblr}{
  width=\columnwidth,
  colsep=3pt, rowsep=2pt,
  colspec={X[l,0.9] X[l,1.0] X[c,0.9] X[c,0.9] X[c,1.2] X[j,1.3] X[c,0.8]},
  rulesep=0pt,
  stretch=0.95,
  hline{1,Z} = {1pt},
}
\midrule
\cite{Misra2000Fluid-basedRED} &
  SDE fluid queuing model &
  Full network scenarios &
  Flow-level (temporal) &
  TCP (Generic w/ RED) &
  Flow throughput, average RTT (but evaluation only evaluates queue-length) &
  Simulated data evaluation \\ \midrule 
\cite{Liu2003FluidNetworks} &
  SDE fluid queuing model &
  Full network scenarios &
  Flow-level (temporal) &
  TCP (Sack, Reno, and NewReno) &
  Flow throughput, average RTT (but evaluation only evaluates queue-length and window size) &
  Simulated data evaluation \\ \midrule 
multi-AIMD \cite{Baccelli2003FlowNetworks} &
  Fluid/ discrete hybrid queuing model &
  Full network scenarios &
  Flow-level (temporal) &
  TCP and UDP &
  Flow throughput, average RTT &
  Simulated data evaluation \\ \midrule 
\cite{Bohacek2003ANetworks}, \cite{Lee2007ModelingSystems} &
  Fluid/ discrete hybrid queuing model &
  Full network scenarios &
  Flow-level (temporal) &
  TCP and UDP &
  Average flow throughput and RTT (only drop rate and window size retain a temporal component) &
  Simulated data evaluation \\ \midrule 
\cite{Marsan2004UsingNetworks} &
  PDE fluid queuing model &
  Traffic-matrix scenarios &
  Flow-level (temporal) &
  TCP (Generic) &
  Average flow throughput, FCT and loss rate &
  Simulated data evaluation \\ \midrule 
\cite{Baccelli2009StochasticTCP} &
  PDE fluid queuing model &
  Traffic-matrix scenarios &
  Flow-level (temporal) &
  TCP (Scalable TCP) &
  Flow throughput &
  Simulated data evaluation \\ \midrule 
\cite{Carofiglio2010OnPerformance} &
  ODE fluid queuing model &
  Full network scenarios &
  Flow-level (temporal) &
  UDP$^{\mathrm{2}}$ and TCP (Reno) &
  Flow RTT, throughput and loss rate &
  Simulated data evaluation \\ \midrule 
\end{tblr}
\begin{tabular}{m{\columnwidth}}
    $^{\mathrm{2}}$ UDP only as supported as background traffic. \\
\end{tabular}
\end{table*}

\begin{table*}[!h] 
\centering
\small
\begin{tblr}{
  width=\columnwidth,
  colsep=3pt, rowsep=2pt,
  colspec={X[l,0.9] X[l,1.0] X[c,0.9] X[c,0.9] X[c,1.2] X[j,1.3] X[c,0.8]},
  rulesep=0pt,
  stretch=0.95,
  hline{1,Z} = {1pt},
}
\midrule 
\cite{Czachorski2015QueueingAnalysis} &
  Discrete and fluid queuing model &
  Traffic-matrix scenarios &
  Flow-level (temporal) &
  TCP &
  Flow packet delay (evaluation focuses in average queue lengths) &
  Simulated data evaluation \\ \midrule 
TFA \cite{Cruz1991AIsolation, Cruz1991AAnalysis} &
  ADNC &
  Full network scenarios &
  Flow-level (temporal) &
  Non-specific &
  Worst-bound packet delay &
  Analytical evaluation \\ \midrule
D-BIND \cite{Knightly1997D-BIND:Traffic} &
  ADNC &
  Full network scenarios &
  Flow-level (temporal) &
  Non-specific &
  Worst-bound packet delay &
  Real data evaluation \\ \midrule 
\cite{Agrawal1999PerformanceProtocols} &
  ADNC &
  Full network scenarios &
  Flow-level (temporal) &
  UDP and traffic with throughput guarantees &
  Worst-bound packet delay &
  Analytical evaluation \\ \midrule 
SFA \cite{2001NetworkCalculus} &
  ADNC &
  Full network scenarios &
  Flow-level (temporal) &
  Non-specific &
  Worst-bound packet delay &
  Analytical evaluation \\ \midrule 
PMOO\cite{Fidler2003ExtendingScheduling} &
  ADNC &
  Full network scenarios &
  Flow-level (temporal) &
  Non-specific &
  Worst-bound packet delay &
  Simulated data evaluation \\ \midrule 
\cite{Lampka2016AchievingDomains} &
  ADNC &
  Full network scenarios &
  Flow-level (temporal) &
  Non-specific &
  Worst-bound packet delay &
  Analytical evaluation \\ \midrule 
TMA \cite{Bondorf2017QualityAnalysis} &
  ADNC &
  Full network scenarios &
  Flow-level (temporal) &
  Non-specific &
  Worst-bound packet delay &
  Simulated data evaluation \\ \midrule 
\cite{Schmitt2008DelayLurch...}, &
  ODNC &
  Full network scenarios &
  Flow-level (temporal) &
  Non-specific &
  Worst-bound packet delay &
  Analytical and Simulated data evaluation \\ \midrule 
\cite{Bouillard2016TightNetworks} &
  ODNC &
  Full network scenarios &
  Flow-level (temporal) &
  Non-specific &
  Worst-bound packet delay &
  Simulated data evaluation \\ \midrule  
\end{tblr}
\end{table*}

\begin{table*}[!h] 
\centering
\small
\begin{tblr}{
  width=\columnwidth,
  colsep=3pt, rowsep=2pt,
  colspec={X[l,0.9] X[l,1.0] X[c,0.9] X[c,0.9] X[c,1.2] X[j,1.3] X[c,0.8]},
  rulesep=0pt,
  stretch=0.95,
  hline{1,Z} = {1pt},
}
\midrule 
\cite{Kiefer2010SearchingNetworks} &
  ODNC &
  Full network scenarios &
  Flow-level (temporal) &
  Non-specific &
  Worst-bound packet delay &
  Simulated data evaluation \\ \midrule 
\cite{Cheng-ShangChang1994StabilityNetworks} &
  SNC &
  Traffic-matrix scenarios &
  Flow-level (temporal) &
  Non-specific &
  Stochastic worst-bound packet delay &
  Analytical evaluation \\ \midrule 
\cite{Jiang2006ACalculus} &
  SNC &
  Traffic-matrix scenarios &
  Flow-level (temporal) &
  Non-specific &
  Stochastic worst-bound packet delay &
  Analytical evaluation \\ \midrule 
\cite{Angrishi2013} &
  SNC &
  Traffic-matrix scenarios &
  Flow-level (temporal) &
  Non-specific &
  Stochastic worst-bound packet delay &
  Analytical evaluation \\ \midrule 
\cite{Lakshman1997TheLoss} &
  Analytical model from RFC definitions &
  Single-flow scenarios &
  Flow-level (temporal) &
  TCP (generic with phase effects and RED) &
  Flow throughput &
  Simulated data evaluation \\ \midrule 
\cite{Padhye2000ModelingValidation} &
  Stochastic analytical model from observations &
  Single-flow scenarios &
  Flow-level (temporal) &
  TCP (Reno) &
  Flow packet rate and throughput &
  Real data evaluation \\ \midrule 
\cite{87140} &
  Utility maximization model &
  Traffic-matrix scenarios &
  Flow-level &
  TCP &
  Flow throughput &
  Analytical evaluation \\ \midrule 
\cite{Kelly1998RateStability} &
  Utility maximization model &
  Traffic-matrix scenarios &
  Flow-level &
  TCP &
  Flow throughput &
  Analytical evaluation \\ \midrule 
\end{tblr}
\caption{Summary of analytical network performance models}
\label{tab:analytical}
\end{table*}

\end{landscape}

Alternatively, relaxing these assumptions will result in more complex models, like the G/G/1 queue, which assumes general distributions for both packet interarrival and service times, or the modified M/G/$\infty$ queue in~\cite{Poon2001} for modeling video traffic.
By not being bound by such assumptions, the resulting model is harder to solve, and may not even have an exact solution \cite{whitt1993approximations}. Usually, approximate solutions are extracted from solvers such as Buzen's convolution algorithm~\cite{chandy1972analysis}, mean value analysis~\cite{10.1145/322186.322195}, Local Balance Algorithm for Normalizing Constants and Coalesce Computation of Normalizing Constants \cite{Chandy1980ComputationalNetworks}. A recent proposal is that of~\cite{Fiorini2025}, where the authors use nonstandard analysis for analyzing GI/G/1 queues.

Hence, it is also common to build more complete models using an ensemble of simpler queue models.
An example can be found in \cite{Garetto2001AFlows, Garetto2004ClosedFlows}, where a model of TCP Tahoe was built out of a graph made out of M/M/$\infty$ queues (and later M/G/$\infty$, assuming a generalized service time distribution), each representing a state of the protocol and which the packets traversed. Their model also covers link interfaces through an M/M/1/B queue, but only considers line topologies in their evaluation. Another example can be found in \cite{Yu2010ATraffic}, where the authors propose a model capturing bursty and phase-type traffic in a multistage switch network. It explains how these traffic profiles are aggregated and uses adapted M/M/1 queues to approximate the more complex arrival distributions.

QT models are further limited by only working through distributions. First, it only allows models to predict aggregate (e.g., average) performance measures, such as the average packet delay. Second, no matter how expressive the distribution is, it will always be less descriptive than working with individual traffic packets, as is done in DES. Next, they tend to model only traffic flows, independently of the underlying network hardware. The latter is abstracted into ``servers" that delay packets by a given amount, according to a constant or exponential distribution, depending on the queue model.

We also note that early queuing models focused on highly specific scenarios, typically modeling single flows without accounting for cross-traffic from other TCP connections or background traffic. That being said, this limitation has been addressed in the SotA \cite{Garetto2001AFlows, Garetto2004ClosedFlows, Yu2010ATraffic}, incrementally making them viable in a wider range of scenarios. That being said, queuing models for TCP implementations are susceptible to becoming outdated as networks evolve and the TCP flavor they support becomes deprecated. In spite of this, QT models remain a useful tool for network operators to quickly obtain a general and approximate description of the expected behavior.

\subsection{Fluid Models}

Fluid models (a.k.a. liquid models) are a subtype of QT models. Unlike standard (discrete) QT models, which account for data transmission in indivisible packets, fluid models treat data as a continuous, infinitely divisible flow. This allows for more expressive models to be developed based on systems of Ordinary, Stochastic, and, later on, Partial Differential Equations (ODE, SDE, and PDE, respectively).

Since plenty of ODE and PDE solvers exist, due to their prevalence in other fields, fluid models tend to be computationally inexpensive to solve. Furthermore, they are more expressive, as their description by sets of differential questions also allows authors to include a temporal component in their metrics predictions, unlike discrete models that focus on aggregated (e.g., average) results. 
Being able to formulate the network state through a series of equations, akin to other fields like thermodynamics and electrical systems, is also appealing.

However, ODEs also introduce important limitations. By themselves, these systems introduce a degree of error through approximation —solvers based on numerical solutions, for example, will inevitably introduce a degree of error due to their approximate approach. These inaccuracies are much more prevalent in computer networks due to their discrete nature: digital information is measured in bits and sent through indivisible packets; decisions taken by transport protocol algorithms, like TCP, are made at a resolution of these individual packets. Hence, any formulation that deals with the network as a continuous system will be limited due to this fundamental difference.

Nonetheless, its advantages prevailed, and fluid models proliferated in the SotA between 2000 and 2010.
For example, in \cite{Bonald1999ComparisonFairness}, the authors implemented a fluid model of both TCP Reno and TCP Vegas to compare the two.
The comparison, however, is limited to the flow control section of both flavors.
In AIMD~\cite{Baccelli2002AIMDTraffic}, the authors propose a generic TCP model, including a steady-state solution to extract the flow's throughput and mean stationary inter-congestion time. However, like other generic TCP models, it does not cover advances and differences between TCP implementations. In \cite{Bohacek2003ATransmission}, the authors propose a generic TCP model to predict performance metrics relevant to video transmission (packet RTTs and loss rates). The model assumes that packet loss can be described as a Poisson Process.

Next, in~\cite{Misra2000Fluid-basedRED}, the authors define a system of SDE, later converted to a system of ODE, to model Random Early Detection (RED), a mechanism of Active Queue Management (AQM). The model can be expanded from modeling a single router to a network with multiple TCP flows, and scales to larger instances through flow aggregation. However, the traffic matrix it uses ignores the order of visited routers and assumes a generic version of TCP. Shortcomings of this model are later addressed in \cite{Liu2003FluidNetworks}: it expands the RED model to include other types of AQM policies, specific versions of TCP (SACK, Reno, and NewReno), and expands the routing information from the network topology considered by it. The model also introduces pruning of irrelevant elements (e.g., non-congested queues) to reduce the number of variables and improve computational performance.

There are also examples of models that attempt to combine both discrete and fluid queuing models. In multi-AIMD~\cite{Baccelli2003FlowNetworks}, the authors expand the AIMD model to support the presence of both TCP and UDP flows. It considers a fluid model that feeds into an M/M/1/B (discrete) queue model for the network devices according to the network topology. In~\cite{Bohacek2003ANetworks, Lee2007ModelingSystems}, the authors follow a different approach, where flows are described through a discrete state (i.e., empty queues, non-congested flow, and congested flow), each with its own system of SDE. It allows modeling a more complex set of behaviors than a single SDE. The model supports different TCP flavors and UDP, albeit the latter following a specific On/Off interarrival traffic distribution. Importantly, the evaluation only covers predictions of the average flow's throughput and RTTs over the scenario, and the papers do not specify how to compute the time-aware values.

One of the first examples of a model that uses a system of PDEs can be found in \cite{Marsan2004UsingNetworks}. The model exploits the increased expressiveness of fluid models when expressing flow behavior over time, to better deal with ``mice" flows. These are short flows whose completion time is more dependent on propagation delays than transmission delays. The model considers a generic TCP version. It still relies on assumptions (e.g., mice flow lengths are exponentially distributed) and does not address the underlying topology. In the evaluation, despite including the temporal component in its reasoning, they only consider average values for the throughput and FCT. Later, in \cite{Baccelli2009StochasticTCP}, the authors define a model using PDEs to model Scalable TCP. While this model supports multiple flows, it assumes constant RTT and continues to disregard the network topology.

Later on, we have models that cover more complex network scenarios. For example, in~\cite{Carofiglio2010OnPerformance} the authors develop an ODE model for considering queuing policies other than FIFO. This includes fair queuing, longest queue first, and shortest queue first. The model also supports multiple TCP flows and background UDP traffic. The model can predict the TCP flow's throughput and buffer occupancy, and the UDP's loss rate. However, the model is adjusted for TCP Reno, specifically during the congestion avoidance phase. Another example is in \cite{Czachorski2015QueueingAnalysis}, where the authors modify both discrete and fluid queuing models to capture transient states of TCP flows. While the obtained fluid model is less accurate, its lower computational cost allows it to be applicable on larger networks, unlike the discrete model.

\subsection{Network Calculus}


Network Calculus (NC) is a set of techniques used to study network performance introduced in~\cite{Cruz1991AIsolation, Cruz1991AAnalysis}. Unlike QT, which describes network traffic through traffic distributions, NC describes traffic through arrival and service curves. These are then used to generate worst-case bounds for the flow's performance metrics over time. In~\cite{Cruz1991AIsolation}, the authors defined network elements, like delay lines, buffers, and regulators. Then, in~\cite{Cruz1991AAnalysis}, the authors show how these elements can be combined to define a network model.
Briefly, NC became popular as a middle point between QT and DES, offering inexpensive predictions, relying on fewer traffic assumptions, and with an integrated temporal component

NC models, however, are not flawless. While NC ensured a correct worst-case bound, originally these were too permissive for the prediction to be useful. Conversely, tighter bounds may result in increased model complexity and computational cost. Due to this, SoTA NC models attempt to push the Pareto front by pursuing a better balance between tight bounds and the cost of obtaining them \cite{Bondorf2017QualityAnalysis}. Ultimately, NC models can be subdivided into three categories: Algebraic-based Discrete Network Calculus (ADNC), Optimization-based Discrete Network Calculus (ODNC), and Stochastic Network Calculus (SNC).

Finally, a significant limitation shared across all NC models is that it can only work in feedforward networks —i.e., where routing paths do not result in cycles. This aspect was studied in \cite{Charny2000DelayScheduling}, where the authors tried to define a delay bound for generalized topologies, but only found it possible when link utilization is extremely low. Otherwise, the packet delay for a given flow may be influenced by flows that do not share a queue within their routing paths or have been completed before the given flow has even started. As the authors state, a delay bound may not even exist beyond a large enough link utilization.
This can be addressed by turning general network topologies into functionally feedforward ones. For example, networks can be shaped into spanning trees by removing links from consideration.
Alternatively, in~\cite{Starobinski2003ApplicationTurn-prohibition} they propose modifying the flow's routing paths through turn-prohibition. While the latter is less drastic, both decrease flow throughput. Empirical evaluation suggests that this effect worsens in larger topologies.

\subsubsection{Algebraic-based Discrete Network Calculus (ADNC)}
It is the approach followed by the original proposal \cite{Cruz1991AIsolation, Cruz1991AAnalysis}, where network elements are defined and can be chained. The delay bounds are solved algebraically through the use of (min, $+$) and (max, $+$) algebra \cite{2001NetworkCalculus}. Later on, the original proposal was referred to as Total Flow Analysis (TFA). One of the first improvements was using piece-wise functions to replace the linear arrival and service curves, as used in \cite{Knightly1997D-BIND:Traffic}, resulting in more expressive and tighter definitions. \cite{Agrawal1999PerformanceProtocols} also extends the original proposal by supporting flows with flow control —i.e., flows that reserve network resources for performance guarantees. Later in~\cite{2001NetworkCalculus}, the authors proved that TFA's delay bound overestimates the impact of flow bursts at each node in a flow's path. This can be addressed by defining and exploiting the concatenation property (known as the Pay Bursts Only Once Principle or PBOO), resulting in Separate Flow Analysis (SFA).

The PBOO principle was extended in \cite{Fidler2003ExtendingScheduling} to make it applicable to apply NC for multiplexing flows in the same node, renamed as Pay Multiplexing Only Once (PMOO). In theory, PMOO also allowed for considering arbitrary multiplexing of flows, unlike previous approaches, which assumed FIFO multiplexing. However in \cite{Schmitt2008DelayLurch...} shows that arbitrary multiplexing may result in looser bounds than those obtained with SFA. Next, the authors at \cite{Lampka2016AchievingDomains} propose extending the PMOO model using bounded arrival curves, making the model more efficient with the aim of unlocking the use of more accurate yet complex function definitions for the arrival and service curves. Finally, in~\cite{Bondorf2017QualityAnalysis} the authors compare both ADNC and ODNC alternatives in the SotA, and offer an improved version of ADNC, named Tandem Matching Analysis (TMA). It considers all possible (min, $+$) operations during the network decomposition to obtain the least pessimistic delay bound. While this approach results in an exponential number of decompositions according to the network size, it is optimized to keep computational costs akin to similar alternatives, as proved by their evaluation.

\subsubsection{Optimization-based Discrete Network Calculus (ODNC)}
These models find the delay bound for a given flow in the network by solving a Linear Programming Problem (LPP). The LPP is formulated to identify the worst possible delay according to a set of temporal, spatial, and service constraints. An early example is introduced at \cite{Schmitt2008DelayLurch...} as an alternative the PMOO model. While more accurate, especially when facing complex scenarios like non-FIFO flow multiplexing, this approach suffers from an extremely high computational cost. This is because solving the delay bound for a given flow is an NP-hard problem, as both the number of LPPs to solve and constraints in each of them increase exponentially with the network size \cite{Bouillard2016TightNetworks}.

Hence, authors have looked at heuristics or scenarios where the computational cost is more bounded. In the same paper~\cite{Bouillard2016TightNetworks}, the authors find that when considering tandem networks —graphs that can be described as a directed path with no shortcuts— the computational cost decreases from exponential to polynomial. They also prove a heuristic where a universal service curve can be found for all flows in the network, further reducing the computational cost. Another heuristic is proposed in \cite{Kiefer2010SearchingNetworks} where a combination of Monte-Carlo and Direct Search was used to reduce the cost of solving the LPP. However, an analysis done by~\cite{Bondorf2017QualityAnalysis} shows that these heuristics are insufficient to make solving the problems feasible for large networks.

\subsubsection{Stochastic Network Calculus (SNC)}
While traditional (deterministic) NC offers a strict delay bound, SNC instead defines delay bounds that can be incorrect with a small probability of error~\cite{Fidler2015ACalculus}. The earliest paper proposing this approach can be found in \cite{Cheng-ShangChang1994StabilityNetworks}, where they use bounds for the moment generating functions of random variables. While there have been several attempts to define a complete SNC \cite{10.5555/1434872}, the authors at \cite{Jiang2006ACalculus} are the first to aggregate formulations from previous research to build one that satisfied a set of listed properties (e.g., service guarantees, per-flow service). However, later work by \cite{Ciucu2012PerspectivesCalculus} showed that such a model is not purely stochastic and struggles in aspects such as capturing statistical flow multiplexing. In~\cite{Angrishi2013}, the author proposes expanding existing SNC by estimating the effective network bandwidth to obtain better delay bounds. Ultimately, the biggest drawback of SNC is its relatively recent development, making it less explored and developed than its deterministic counterparts.

\subsection{Other Analytical Models}

Analytical models that are not based on either QT or NC are generally rare, as this implies losing the accumulated knowledge from previous research. Nonetheless, sometimes models may be built based on other principles, or the authors wish to avoid the weaknesses present in these fields.

A common source of alternative analytical models is building ones based on the empirical observations, or, in the case of well-defined protocols like TCP, on the RFC definitions. In~\cite{Lakshman1997TheLoss}, the authors implement an analytical model for TCP with forward acknowledgments using Rate-Halving. This model was later evaluated considering phase effects and RED. However, the evaluated topologies were small, generally single-link topologies, and the Internet's topology was simplified to a single link with losses between the routers. Next, in~\cite{Padhye2000ModelingValidation}, the authors build a stochastic model for TCP Reno. It processes flows into rounds according to the congestion window. Unfortunately, it relies on several assumptions for the model to be solved through a closed-form solution: losses are independent of those in different rounds, and RTT is assumed to be independent of the window size. The model also does not cover Reno's fast-recovery algorithm or the small differences across implementations.

Alternatively, there has been research applying game theory to model congestion control algorithms. This idea was introduced in~\cite{87140}, where the congestion algorithm was tasked to find a Nash equilibrium where performance across all present flows is maximized. The authors propose using the Nash arbitration scheme to predict the throughput allocation for each flow while proving the existence of an optimal allocation. Inspired by it, in~\cite{Kelly1998RateStability} proposes two models. The first, an ideal model, relies on knowing the ``utility" obtained by each flow at a given throughput, which is unknown beforehand. The second model simplifies the first, being decomposed into a solvable dual problem.
A strength of this approach is that models considered link capacities in their reasoning, which was rare for analytical models at the time. Conversely, this approach had two main drawbacks. First, in practice, congestion algorithms like TCP may not necessarily allocate throughput solely by maximizing bandwidth allocation. Second, the proposed models are generally too complex to be applied to real networks, while their simplified forms introduce inaccuracies. 

\subsection{Summary}

In summary, analytical models aim to be a cost-effective alternative to DES. The first models are based on Queueing Theory, which uses queues or systems of queues to describe both network devices and protocols~\cite{Jackson1963Jobshop-likeSystems, Kobayashi1977QueueingAnalysis}. However, simpler queue models lack accuracy due to oversimplistic assumptions, while more complex models are no longer cost-effective to solve. Fluid models are also based on QT, but allow flow payloads to be infinitely divisible. While this contradicts how traffic packets behave in reality, it allows models to be formulated through a system of differential equations, reinforcing their cost-effectiveness while incorporating the temporal component in their prediction~\cite{Bonald1999ComparisonFairness}.

In response to QT's limitations, Network Calculus was developed as an alternative for network modeling~\cite{Cruz1991AIsolation, Cruz1991AAnalysis}. Unlike queuing models, NC models included the temporal component and predicted performance bounds rather than direct estimations. However, like QT models, NC models are also split between inaccurate predictions (i.e., loose bounds) or prohibitively computational costs. Finally, while there are analytical models that do not fall under any of these fields, their use tends to be more sporadic and more specialized (e.g., using Game Theory to understand bandwidth allocation in congestion-controlled traffic~\cite{87140, Kelly1998RateStability}).



\section{Machine Learning Models}
\label{sec:ml_models}
This section focuses on network models that predict network performance behavior through the use of ML models. Discussed models are summarized in Table~\ref{tab:ml}.

\subsection{``Shallow" ML Models}

The earliest attempt to use ML to predict network performance that we identified was in \cite{He2005OnThroughput}. In it, the authors considered both ``formula-based" (analytical) and ``history-based" (temporal regression) models when predicting TCP throughput. They proved analytical models' limitations when dealing with saturated traffic, while even simple regression models like moving average (MA) or exponentially weighted MA (EWMA), worked fairly well. Still, the purposes of these models were not to build a tool, but to evaluate the feasibility of ML models.

\begin{landscape}

\begin{table*}[!hp] 
\centering
\small
\begin{tblr}{
  width=\columnwidth,
  colsep=3pt, rowsep=2pt,
  colspec={X[l,0.9] X[l,1.0] X[c,0.9] X[c,0.9] X[c,1.2] X[j,1.3] X[c,0.8]},
  rulesep=0pt,
  stretch=0.95,
  row{1} = {font=\bfseries},
  hline{1,Z} = {1pt},
  hline{2}   = {0.6pt},
}
Model &
  Type &
  Input scope &
  Output scope &
  Traffic type &
  Performance metrics &
  Evaluation \\ \midrule
\cite{He2005OnThroughput} &
  MA, EWMA, Holt-Winters &
  Single-flow scenarios &
  Flow-level (temporal) &
  TCP (generic) &
  Flow throughput &
  Testbed data \\ \midrule
PathPerf \cite{Mirza2010APrediction} &
  SVR &
  Single-flow scenarios &
  Flow-level (temporal) &
  TCP (generic) &
  Flow throughput &
  Tesbed data and real data \\ \midrule
WISE \cite{Tariq2013AnsweringExperience} &
  Causal graph &
  Single-flow scenarios &
  Flow-level &
  HTTP connections &
  Network response time &
  Real data \\ \midrule
\cite{Helm2022Flow-levelTheory} &
  EVT &
  Multi-flow scenarios &
  Flow-level &
  UDP &
  50th, 75th, 90th, 99th, 99.9th, 99.99th, 99.999th and 100th percentiles for packet delay and jitter &
  Testbed data \\ \midrule
CLAAP \cite{Monaco2025Real-timeApplications} &
  RBFNN &
  Single-flow scenarios &
  Flow-level (temporal) &
  UDP &
  Flow latency &
  Real data \\ \midrule
\cite{Mestres2018UnderstandingNetworks} &
  DNN &
  Multi-flow scenarios &
  Flow-level &
  UDP &
  Average packet delay &
  Simulated data \\ \midrule
\cite{Krasniqi2020End-to-endSampling} &
  DNN, RF &
  Multi-flow scenarios &
  Flow-level &
  UDP and TCP &
  Average packet delay &
  Simulated data \\ \midrule
RouteNet \cite{Rusek2019UnveilingSDN, Suarez-Varela2019ChallengingModeling, Rusek2020RouteNet:SDN} &
  Custom-designed MPNN &
  Full network scenarios &
  Flow-level &
  UDP &
  Average packet delay and jitter &
  Simulated data \\ \midrule
\cite{Badia-Sampera2019TowardsNetworks} &
  Custom-designed MPNN &
  Full network scenarios &
  Flow-level &
  UDP &
  Average packet delay and jitter &
  Simulated data \\ \midrule

\end{tblr}
\end{table*}

\begin{table*}[!h] 
\centering
\small
\begin{tblr}{
  width=\columnwidth,
  colsep=3pt, rowsep=2pt,
  colspec={X[l,0.9] X[l,1.0] X[c,0.9] X[c,0.9] X[c,1.2] X[j,1.3] X[c,0.8]},
  rulesep=0pt,
  stretch=0.95,
  hline{1,Z} = {1pt},
}
\midrule
RouteNet-Erlang \cite{Ferriol-Galmes2022RouteNet-Erlang:Evaluation} &
  Custom-designed MPNN &
  Full network scenarios &
  Flow-level &
  UDP &
  Average packet delay and jitter, loss rate &
  Simulated data \\ \midrule
RouteNet-Fermi \cite{Ferriol-Galmes2023RouteNet-Fermi:Networks} &
  Custom-designed MPNN &
  Full network scenarios &
  Flow-level &
  UDP &
  Average packet delay and jitter, loss rate &
  Simulated data and testbed data \\ \midrule
RouteNet-Gauss \cite{Guemes-Palau2025RouteNet-Gauss:Learning} &
  Custom-designed MPNN with temporal component &
  Full network scenarios &
  Flow-level (temporal) &
  UDP &
  [Average, median, and 90th, 95th, and 99th percentile] packet [delay, jitter] &
  Testbed data \\ \midrule
\cite{Dhamala2024PerformanceMechanism} &
  Custom-designed MPNN with graph attention &
  Full network scenarios &
  Flow-level &
  UDP &
  Average packet delay &
  Simulated data \\ \midrule
\cite{ClaudioModesto2024DelayDataset} &
  Custom-designed MPNN with graph attention &
  Full network scenarios &
  Flow-level &
  UDP &
  Average packet delay &
  Simulated data \\ \midrule
\cite{KaanAykurt2024DigitalData} &
  Custom-designed MPNN &
  Full network scenarios &
  Flow-level &
  UDP &
  Average packet delay &
  Testbed data \\ \midrule
\cite{Guemes-Palau2024Wavelet-EnhancedModeling} &
  Custom-designed MPNN &
  Full network scenarios &
  Flow-level &
  UDP &
  Average packet delay &
  Testbed data \\ \midrule
\end{tblr}
\end{table*}

\begin{table*}[!h] 
\centering
\small
\begin{tblr}{
  width=\columnwidth,
  colsep=3pt, rowsep=2pt,
  colspec={X[l,0.9] X[l,1.0] X[c,0.9] X[c,0.9] X[c,1.2] X[j,1.3] X[c,0.8]},
  rulesep=0pt,
  stretch=0.95,
  hline{1,Z} = {1pt},
}
\midrule
DeepComNet \cite{Geyer2019DeepComNet:Learning} &
  GGNN \cite{li2016gated} &
  Full network scenarios &
  Flow-level &
  UDP and TCP &
  TCP flow throughput, UDP average packet delay &
  Simulated data \\ \midrule
\cite{Jaeger2022ModelingNetworks} &
  GGNN \cite{li2016gated} &
  Full network scenarios &
  Flow-level &
  TCP (Reno, Cubic, Bic, Illinois, Veno, Vegas and Ledbat) &
  Flow RTT and throughput &
  Simulated data \\ \midrule
\cite{Suzuki2020OnLearning} &
  GCN \cite{kipf2016semi} &
  Full network scenarios &
  Flow-level &
  UDP and TCP &
  Average packet delay &
  Simulated data \\ \midrule
EAGLE \cite{Liu2023EAGLE:Analysis} &
  Custom-designed MPNN &
  Full network scenarios &
  Flow-level &
  UDP and TCP &
  Average packet delay and loss rate &
  Simulated data \\ \midrule
xNet \cite{Huang2024XNet:Networks} &
  Custom-designed MPNN with temporal component &
  Full network scenarios &
  Flow-level (temporal) &
  UDP and TCP (DTCP) &
  Average packet delay, flow throughput and FCT &
  Simulated data \\ \midrule
GLANCE \cite{Li2024GLANCE:Networks} &
  Custom-designed MPNN &
  Full network scenarios &
  Flow-level &
  UDP &
  Average packet delay and jitter, flow throughput, loss rate &
  Simulated data \\ \midrule
FlowSeer \cite{Du2024FlowSeer:Level} &
  Ensemble of GCN \cite{kipf2016semi}, DNN with attention and DNN &
  Full network scenarios &
  Flow-level &
  UDP &
  Average packet delay &
  Simulated data \\ \midrule
m4 \cite{Li2025M4:Simulator} &
  GraphSage \cite{10.5555/3294771.3294869} + GRU \cite{cho2014learningphraserepresentationsusing} &
  Full network scenarios &
  Flow-level (temporal) &
  TCP (DCTCP, TIMELY, DCQCN) &
  [Average, 90th percentile] flow FCT and throughput &
  Simulated data \\ \midrule
\end{tblr}
\end{table*}

\begin{table*}[!h] 
\centering
\small
\begin{tblr}{
  width=\columnwidth,
  colsep=3pt, rowsep=2pt,
  colspec={X[l,0.9] X[l,1.0] X[c,0.9] X[c,0.9] X[c,1.2] X[j,1.3] X[c,0.8]},
  rulesep=0pt,
  stretch=0.95,
  hline{1,Z} = {1pt},
}
\midrule
xWeaver \cite{Wang2018NeuralLearning} &
  CNN &
  Full network scenarios &
  Flow-level &
  UDP &
  Flow FCT and throughput &
  Simulated data and testbed data \\ \midrule
Deep-Q \cite{Xiao2018Deep-Q:Network} &
  CVAE \cite{kingma2022autoencodingvariationalbayes} + LSTM \cite{10.1162/neco.1997.9.8.1735} &
  Multi-flow scenarios &
  Flow-level (temporal) &
  UDP &
  Packet delay and loss rates &
  Testbed data \\ \midrule
\cite{Dietmuller2022AGeneralization} &
  Transformer \cite{vaswani2023attentionneed} &
  Single-flow scenarios &
  Packet-level &
  UDP &
  Packet delay &
  Simulated data \\ \midrule
DeepQueueNet \cite{Yang2022DeepQueueNet:Visibility} &
  Transformer \cite{vaswani2023attentionneed} + BiLSTM &
  Full network scenarios &
  Packet-level &
  UDP &
  Full packet path information &
  Simulated data \\ \midrule
\cite{Hattori2024MetaMetrics, Hattori2025MetaMetrics} &
  NP \cite{garnelo2018neuralprocesses} &
  Multi-flow scenarios &
  Flow-level &
  Non-specific &
  Average packet delay, flow throughput and loss rates &
  Testbed data \\ \midrule
\end{tblr}
\caption{Summary of ML network performance models}
\label{tab:ml}
\end{table*}

\end{landscape}

Later in PathPerf~\cite{Mirza2010APrediction}, the authors used a Support Vector Regression (SVR) model to predict TCP throughput. It was trained in a testbed using real network data and diverse routing paths, showing competitive performance relative to analytical models. SVRs, a popular architecture at the time, are able to learn from limited amounts of data, which is extremely useful considering the cost and difficulties of capturing real-world data for training. However, this architecture still had several limitations: the model could only consider one TCP flow at a time, could only predict average metrics, and the model had to be retrained for every change to the flow's path. 

Another example of an early performance prediction tool is WISE~\cite{Tariq2013AnsweringExperience}. WISE sits on the edge between performance prediction and network validation: it builds a causal graph from captured traffic traces that focus on HTTP connections and the identified variables, such as the number of packets transmitted, their size, and estimates network bandwidth. A causal graph is a statistical model that defines how variables influence each other in a non-cyclical graph. WISE focused on predicting the HTTP connection's response time, both from the network and browsers. The causal graph allows the user to query different values for the specified variables, hence allowing the evaluation of ``what-if" scenarios to understand their impact on the response time.
In the evaluation, which consisted of captured data from Google's network, it proved to be accurate in predicting both options of the response time. Unlike other ML models, it does not need training. Instead, it learns and builds the causal graph during inference, making the method versatile but slow.

Nowadays, ``shallow" ML models have fallen out of favor against their DL counterparts, as the latter tend to offer more accurate results due to their more complete architecture. However, ``shallow" ML models are sometimes desired due to their efficiency, being able to be trained with few samples and at a low computational cost. A recent example of this is in \cite{Helm2022Flow-levelTheory}, based on Extreme Value Theory (EVT), a statistical model designed to predict extreme events or worst-case bounds. In this paper, EVT is used to predict upper-bound approximations of metrics like flow latency and jitter. 
The EVT model is highly efficient: it only required to be trained on 5\% of the data available. It was also able to predict scenarios with different topologies and make predictions for the entire network (with multiple-interacting flows) and not only for individual flows. However, the results show that its accuracy for network predictions is significantly lower.
Finally, the EVT model also depends on assumptions like stationary flow latencies.

Another recent example is that of CLAAP~\cite{Monaco2025Real-timeApplications}. CLAAP uses a Radial Basis Function Neural Network (RBFNN) to predict flow latency in online gaming sessions.
It bases its predictions on previous latency measurements, akin to a regression model such as MA and EWMA in~\cite{He2005OnThroughput}. 
While the RBFNN is a shallow neural network, it still outperforms other online regression models in its evaluation. Furthermore, it is designed for quick inference speed, as it is a model meant for online management of online gaming sessions. 
Hence, response time is more critical than creating an accurate yet expensive internal representation of the flow state.

\subsection{Deep Learning and Graph Neural Networks}

The initial attempts to build DL network performance models were tested using Dense Neural Networks (DNNs). In \cite{Mestres2018UnderstandingNetworks}, they evaluate the use of DNNs when predicting the average flow delay. While their results show the potential of DL methods by being computationally inexpensive and capable of modeling non-linear relationships, they noted their inability to generalize to unseen topologies and routings. Also, the only traffic profile of packet arrivals considered was an exponential distribution, a simpler distribution than those faced in real-life scenarios. Later in \cite{Krasniqi2020End-to-endSampling}, the authors compared building a DNN model against a Random Forest (RF) model using a mixture of UDP and TCP traffic of varying complexity. They ultimately showed that the RF models generally outperformed the DNN models. Hence, researchers quickly moved from using DNNs and looked into alternative, better-suited DL architectures.

\subsubsection{RouteNet and Successors}
Among all the possible alternatives, the family of architectures that predominated were Graph Neural Networks (GNNs)~\cite{4700287}. GNNs are designed to take a graph input and exploit the topological information present. Unlike previous ML and analytical models, which struggled to represent the impact of the network topology and flow's routing path on their reasoning, GNNs can properly exploit them, leading to their success. This is exemplified by the RouteNet family of models.

The original one was described and introduced in~\cite{Rusek2019UnveilingSDN}, and was later more thoroughly examined at \cite{Suarez-Varela2019ChallengingModeling, Rusek2020RouteNet:SDN}. RouteNet's architecture is based on the Message Passing Neural Network (MPNN) architecture~\cite{pmlrv70gilmer17a}. It works by building a heterogeneous graph that captures the dependencies between links and flows within a network, based on its topology and routing. 
In principle, using this description allows RouteNet to accurately predict flow delay and jitter in topologies and routing paths unseen during training while maintaining a low computational cost (i.e., milliseconds per network scenario). The RouteNet models are trained and evaluated on samples using DES.

While effective, RouteNet does have important drawbacks. First, further evaluation has shown limitations in its generalization \cite{Happ2022ExploringModeling}. This includes larger topologies and link capacities than those seen during training, but also to different traffic profiles, packet sizes, and inter-packet gap (IPG) time distributions. Furthermore, RouteNet can only predict average measures across the entire scenario, and cannot model transitory behavior. Finally, it does not support key network features, such as modeling TCP traffic, Quality of Service (QoS) requirements, or alternative routing paths.

To address these limitations and improve overall accuracy, RouteNet has since been extended through successive iterations to cover a broader range of scenarios.
The first change was in~\cite{Badia-Sampera2019TowardsNetworks}, in which additional features were added to add support for variable queue sizes.
The first major expansion, however, came with RouteNet-Erlang~\cite{Ferriol-Galmes2022RouteNet-Erlang:Evaluation}. It includes queues in its graph representation of the network alongside links and flows, and its feature extraction is modified to better support inferring over larger topologies than seen during training. It also supports flows parametrized through auto-correlated traffic distributions, based on observations done to internet traffic~\cite{Popoola2017EmpiricalModel}. Overall, this allows for a more complete definition of the network and its traffic, better accuracy when dealing with unseen scenarios during training, and introduces support for multiple queues per port, different queuing policies, supporting QoS requirements, and packet loss predictions. 
The next large iteration is that of RouteNet-Fermi~\cite{Ferriol-Galmes2023RouteNet-Fermi:Networks}. Rather than predicting the average flow delay directly, RouteNet-Fermi instead learns to predict the mean queue occupancy at each step of the flow's routing path. This change in the model's reasoning results in more accurate predictions and better scaling when facing larger networks, in number of elements, traffic intensity, and link capacities. RouteNet-Fermi also expanded the number of supported queuing policies relative to  RouteNet-Erlang. Finally, this iteration of RouteNet was also evaluated with real (testbed) network data, proving its feasibility in more complex traffic scenarios.

The most recent iteration of RouteNet is RouteNet-Gauss~\cite{Guemes-Palau2025RouteNet-Gauss:Learning}.
It introduces a temporal component, splitting traffic scenarios into fixed-time windows, enabling it to model non-stationary traffic and better address the complexities present in real network samples. The window size can be adjusted to balance model expressivity to computational cost. It also removes the need to specify flow distribution parameters as input, allowing it to accurately model non-parametric traffic. Still, RouteNet-Gauss does share some limitations with the original version: specifically, a lack of support for congestion control algorithms and generalization to unseen network protocols and traffic distributions from those seen during training.

Beyond the main line of RouteNet architectures, there have been improvements made by authors other than the original ones. The most common is including an attention mechanism at some point in the message passing phase. In \cite{Dhamala2024PerformanceMechanism}, this is introduced to the original RouteNet, while in \cite{ClaudioModesto2024DelayDataset} it is introduced in the RouteNet-Fermi. The inclusion of graph attention mechanisms is used to improve the expressivity of the message passing phase by allowing nodes to weight the relative importance of their neighbors. This mechanism was introduced in the Graph Attention Network \cite{veličković2018graphattentionnetworks}, which became the SotA general-purpose GNN architecture. We also note that, in \cite{ClaudioModesto2024DelayDataset}, other improvements were introduced, such as using feature selection to refine the feature extraction process.

In \cite{KaanAykurt2024DigitalData}, the authors also expand RouteNet-Fermi by increasing the features used to codify flow information. Specifically, they extract the IPG mean, variance, and several percentiles for each flow. They also propose using the packet loss rate as an input for predicting the flow delay, which, while effective, is impractical in real-life scenarios, as either the packet loss rate is not known during inference, or if it is, the average packet delay is already known as well. 
Finally, in \cite{Guemes-Palau2024Wavelet-EnhancedModeling}, the authors modify RouteNet-Fermi, replacing the use of flow distribution parameters with the raw packet traces and the wavelet decomposition process during flow encoding. This allows for cost-efficient, rich descriptions of non-parametric traffic distributions. The results also suggest the model's ability to perform inference on unseen traffic distributions at a minimal penalty to the prediction error. However, this method requires the packet traces of the flows, which are costly to obtain and not always available.

\subsubsection{Other GNN Models}
While RouteNet has inspired many of the GNN models, there are other alternatives present in the SotA.

Published around the same time as the original RouteNet paper, the authors in~\cite{Geyer2019DeepComNet:Learning} also propose a GNN-based network model. Like RouteNet, the authors also proposed building a graph that captures the dependencies between links and flows. Unlike RouteNet, its architecture is based on Gated GNNs (GGNNs)~\cite{li2016gated} rather than MPNNs. Next, it can model congestion-controlled traffic by representing both the original stream and the ACK messages as separate flows. However, while RouteNet's hypergraph is a heterogeneous graph, with the different network elements being processed differently, the hypergraph in~\cite{Geyer2019DeepComNet:Learning} is a homogeneous graph, and its nodes are only differentiated through a one-hot feature.
Furthermore, its simpler feature extraction results in lower accuracy and generalization to unseen network protocols and traffic intensities.

This approach was later expanded in \cite{Jaeger2022ModelingNetworks}. First, the hypergraph it builds also includes device ports (referred to as interfaces), and path nodes that encode the flow's routings. The encoding of the network elements is also extended to support multiple TCP versions. Overall, it shows lower error rates than their baseline analytical models, but performance varies depending on the TCP version and the routing path length. Furthermore, the TCP version is encoded through one-hot encoding, meaning that it can only be applied to versions covered during training.

Next, in \cite{Suzuki2020OnLearning}, the authors propose using a Graph Convolution Network (GCN)~\cite{kipf2016semi} architecture, rather than an MPNN, to predict traffic delays. Unlike previous networks, they apply the graph to the original network topology instead of building a hypergraph of relationships. Unfortunately, their methodology is not described, and we lack details such as the extracted features. However, we know they evaluated on synthetic 500-node networks and included TCP traffic flows.

In EAGLE~\cite{Liu2023EAGLE:Analysis}, the authors propose a more complex architecture to predict packet delays and loss rates simultaneously. Similar to RouteNet, the network is represented as a hypergraph that describes links, routers, and routing paths. Information between different elements is propagated through random walks across the hypergraph. According to the authors, this technique is parallelizable and effective at extracting neighborhood information, albeit it also introduces several hyperparameters to consider. Once the embeddings are obtained, a multi-head graph attention mechanism is used. While it compares well against their reference queuing model and the original RouteNet, it lacks comparison with other SotA models at the time.

Then, in xNet~\cite{Huang2024XNet:Networks}, they propose another MPNN-based architecture that represents the network through a hypergraph. It considers five types of network elements, more than any other approach: nodes, queues, links, paths, and flows. Paths and flows are later used in the readout to extract the performance delays, such as path packet delay or FCT. A temporal domain is also introduced by splitting the network into fixed-sized windows, achieving this earlier than RouteNet-Gauss. Overall, their evaluation shows better results than models such as Deep-Q~\cite{Xiao2018Deep-Q:Network} and the original RouteNet, but lacks evaluating different traffic distributions or different-sized network topologies.

In GLANCE~\cite{Li2024GLANCE:Networks}, a similar hypergraph definition of the network is proposed, considering path, node, and link embeddings, and an MPNN-like architecture.
Unlike other MPNNs, GLANCE does not reuse parameters between iterations in the message passing phase. This is uncommon, as while it increases the potential model expressivity, in practice, it increases training costs and usually does not result in improved performance.
Also, GLANCE defines an edge graph convolutional layer when updating the node embeddings, combining graph and spectral features. While the latter can capture expressive relationships, they too risk lower generalization to unseen topologies. 
Moreover, GLANCE is the first model to propose the use of transfer learning, specifically, to re-train the model to new tasks quickly. It consists of fixing (a.k.a freezing) the weights of the embeddings and message passing phases, and retraining the readout function from scratch. In the evaluation, GLANCE does improve against RouteNet-Fermi, although this difference is smaller while generalizing to unseen topologies.

Unlike in previous models, FlowSeer~\cite{Du2024FlowSeer:Level} processes topological and flow information separately. The topological information is sent through a GCN, while flow information is processed by a DNN with attention, trained through an encoder-decoder schema. Extracted features are then concatenated and fed to a DNN for the final prediction.
While separating both types of information allows processing each with an appropriate architecture, the model cannot learn how they interact until the final DNN.
In its evaluation, it does show similar or better prediction accuracy than RouteNet-Erlang, but it does not include comparisons of generalization to unseen network topologies.

Next, m4~\cite{Li2025M4:Simulator} proposes the use of a ``flow-level simulator" based on GNNs. Specifically, their solution includes a flow generation module that generates TCP flows according to a user-specified workload.
The model then uses a temporal and spatial component to update the state and predict the flow's performance metrics, specifically FCT and throughput. The temporal component is based on a GRU~\cite{cho2014learningphraserepresentationsusing} model, and the spatial model on a GraphSage~\cite{10.5555/3294771.3294869} model. The spatial component also uses a hypergraph definition, which considers flows and links only.
Unlike RouteNet-Gauss or xNET, its temporal component is based on events (i.e., a flow starting/ending in the network) rather than fixed-sized windows. This minimizes how many times the flow states need to be updated, although it makes it unsuitable for predicting metrics such as packet delay over time. m4 also proposes learning secondary performance metrics during training, such as the flow's remaining data to retransmit, to provide the model with further information and improve its overall accuracy, validated in their evaluation. Finally, m4 cannot model QoS queues.

\subsubsection{Other DL Models}
Due to GNN's advantages, they are the preferred architecture when building network models. However, authors have also experimented with other architectures and approaches.

An early example is that of xWeaver~\cite{Wang2018NeuralLearning}, where authors propose working Convolutional Neural Network (CNN). Specifically, its model uses two separate CNNs to process the adjacency and traffic matrices, whose results are then concatenated and joined in a DNN to predict any relevant performance (in the evaluation, they focus on FCT and throughput). While using a CNN allows for a more effective architecture than a DNN, as corroborated in its evaluation, it still raises some questions. For example, it is not clear if the method is permutation invariant —that is, whether the nodes' order in the traffic matrix impacts the results. Also, similar to FlowSeer, processing separately the topological and traffic information makes it harder for the model to learn cross-interactions.

Another early approach is proposed in Deep-Q~\cite{Xiao2018Deep-Q:Network}, consisting of training a Conditional Variational Auto-Encoder (CVAE) model~\cite{kingma2022autoencodingvariationalbayes}. Both the encoder and decoder segments of the CVAE, implemented through DNNs, use the extracted traffic features as bias. These features are extracted from the traffic matrices using a Long Short-Term Memory (LSTM)~\cite{10.1162/neco.1997.9.8.1735} network. A specialized loss definition is used to train both the LSTM and CVAE. At inference time, only the LSTM and the decoder segment are used to predict the desired performance metrics. While this approach is accurate and has low inference time, it ignores the underlying network topology.

Later, in~\cite{Dietmuller2022AGeneralization}, the authors propose the adoption of the transformer architecture \cite{vaswani2023attentionneed}, popular at the time due to its successful use in LLMs. They argue that having a pre-trained network performance model can be used and later fine-tuned to specific networks. They also defend that transformers, as a powerful sequential model, can be used to generate and use packet embeddings to make packet-level predictions. While they examine some transformer-based network model prototypes, they lack comparisons against SotA models.

A more robust implementation of a network transformer model can be found in DeepQueueNet~\cite{Yang2022DeepQueueNet:Visibility}.
It follows a modular approach: using small, cheap-to-simulate scenarios, it builds a library of DL models, each representing a network device. These are implemented through a Bidirectional LSTM (BiLSTM) and a transformer model. They take as input the sequence of packet arrival times to the device and output the updated sequence after exiting the device. Ultimately, most, if not all, of the devices in the DES are replaced by the DL model counterparts. DeepQueueNet was directly inspired by MimicNet~\cite{Zhang2021MimicNet:Learning}, a hybrid model that replaced aspects of DES simulation with DL, but, unlike it, it is not constrained to a Fat Tree topology.

However, DeepQueueNet's methodology does come with limitations. For example, as noted by their authors, it cannot model the transient state of networks and, because of the batch processing of packets, it cannot support stateful protocols like TCP. Furthermore, because of the batching, packets can be processed out of order; the authors do propose a solution, but it requires repeated inferences by the DL models, increasing computational cost. In addition, both the BiLSTM and transformer architectures are notoriously computationally expensive. As a result, DeepQueueNet does not appear to reduce the amount of computational effort, relative to DES. Instead, due to its implementation through libraries like Tensorflow, it is better suited to run distributively and even in specialized hardware like GPUs. Still, less computationally expensive methods, such as MimicNet, perform inference faster under the same hardware, with the difference scaling with the network topology's size. 

Finally, in \cite{Hattori2024MetaMetrics, Hattori2025MetaMetrics}, the authors propose a two-step process to build a network model by first training it using simulated network data and then efficiently adjusting it using a small dataset of real-world network data. The objective was to utilize transfer learning to allow the model to learn from simulated samples while being effective in real-world scenarios. The model itself was to follow the Neural Processes (NPs)~\cite{garnelo2018neuralprocesses} architecture, a type of DNN that focuses on learning the input's latent features, and eases learning. However, it also shares limitations present in DNNs, like the lack of adaptability to unseen topologies, routing configurations, and traffic profiles during training.

\subsection{Summary}

While ML-based network models have existed since the early 2000s, it was not until the proliferation of deep neural network architectures, and specifically GNNs, that they became so dominant. The flexibility behind these architectures allows for specialized designs for network performance modeling, allowing them to be more accurate than analytical models while performing inference at low inference costs. This makes them especially appealing for active network management, where quick, accurate predictions are required (e.g.,  xWeaver~\cite{Wang2018NeuralLearning}, RouteNet~\cite{Rusek2019UnveilingSDN}, GLANCE~\cite{Li2024GLANCE:Networks}), overtaking analytical models (discussed more in detail in Section~\ref{sec:discussion}). Traditional ML models also remain a viable option, as they are characterized by even cheaper inferences (e.g., CLAAP \cite{Monaco2025Real-timeApplications}).

ML models are not flawless, however. Unlike simulation or analytical models, ML models are constrained to scenarios similar to those seen during training (e.g., network topologies, traffic profiles, network protocols). While this can be mitigated through clever design (e.g., model architecture, how input data is encoded), no covered approach achieves the same degree of generalization as DES. Also, while faster than DES simulators, ML models usually cannot offer the same level of granularity in their predictions. Those that do, like DeepQueueNet~\cite{Yang2022DeepQueueNet:Visibility}, do so at the cost of their low inference costs.

\section{Hybrid Approaches}
\label{sec:hybrid_models}

In this section, we discuss network models that combine two or more of the previously discussed approaches, summarized in Table~\ref{tab:hybrid}. This has the aim of complementing their advantages and minimizing the pitfalls. By nature, these models are the most heterogeneous in the ways they approach the model.

\subsection{Model-tuned Emulation for Performance Modeling}
\label{sec:hybrid_models:emulation}
Network emulators are powerful tools to validate and understand network dynamics. Unlike simulation, which focuses on defining the network state and understanding how it is updated, emulation replicates the network behavior and its devices through software. However, emulation also replicates through software certain behaviors that were originally executed by hardware logic. This distorts the time taken to perform each of these operations, presenting serious limitations when predicting the performance of the networks.

\begin{landscape}

\begin{table*}[!hp] 
\centering
\small
\begin{tblr}{
  width=\columnwidth,
  colsep=3pt, rowsep=2pt,
  colspec={X[l,0.9] X[l,1.1] X[l,1.0] X[c,0.9] X[c,0.9] X[c,1.2] X[j,1.3] X[c,0.8]},
  rulesep=0pt,
  stretch=0.95,
  row{1} = {font=\bfseries},
  hline{1,Z} = {1pt},
  hline{2}   = {0.6pt},
}
Model & Main Type & Secondary Type & Input Scope & Output Scope & Traffic Type & Performance Metrics & Evaluation \\
\midrule
Pantheon \cite{Yan2018Pantheon:Research} &
  Network emulator (Mahimahi \cite{191577}) &
  Bayesian optimization &
  Full network scenarios &
  Packet-level &
  Any &
  Full packet path information &
  Real data \\ \midrule
iBox \cite{Ashok2022Data-DrivenIBox} &
  Network emulator with a queue model &
  Bayesian optimization &
  Full network scenarios &
  Packet-level &
  Any &
  Full packet path information &
  Real data \\ \midrule
Prophet \cite{Zhang2020Prophet:Networks} &
  NUM analytical model~\cite{Kelly1998RateStability} &
  Gradient descent &
  Full network scenarios &
  Flow-level &
  TCP $^{\mathrm{1}}$ &
  Flow throughput &
  Simulated data \\ \midrule
DeepTMA \cite{Geyer2019DeepTMA:Networks, Geyer2020OnCalculus} &
  NC (TMA~\cite{Bondorf2017QualityAnalysis}) &
  GGNN~\cite{li2016gated} &
  Full network scenarios &
  Flow-level (temporal) &
  Non-specific &
  Worst-bound packet delay &
  Simulated data 
  \\ \midrule
\cite{Helm2023PredictingGNNs} &
  GraphSage \cite{10.5555/3294771.3294869} &
  NC (TFA~\cite{Cruz1991AIsolation, Cruz1991AAnalysis}, SFA~\cite{2001NetworkCalculus}) &
  Full network scenarios &
  Flow-level &
  UDP &
  Mean, min, max, 90th and 99th percentile packet delays &
  Testbed data \\ \midrule
QT-RouteNet \cite{deAquinoAfonso2022QT-Routenet:Theory} &
  RouteNet \cite{Rusek2019UnveilingSDN} &
  M/M/1/B queue model &
  Full network scenarios &
  Flow-level &
  UDP &
  Average packet delay &
  Simulated data \\ \midrule
GNNetSlice \cite{Farreras2025GNNetSlice:Networks} &
  RGCN~\cite{schlichtkrull2018modeling} &
  Unspecified QT model &
  Full network scenarios &
  Flow-level &
  UDP &
  Average packet delay, jitter, and loss &
  Simulated data \\ \midrule
\end{tblr}
\begin{tabular}{m{\columnwidth}}
    $^{\mathrm{1}}$ TCP flavors supported by Srikant’s model~\cite{Srikant2004}
    \\
\end{tabular}
\end{table*}

\begin{table*}[!hp] 
\centering
\small
\begin{tblr}{
  width=\columnwidth,
  colsep=3pt, rowsep=2pt,
  colspec={X[l,0.9] X[l,1.1] X[l,1.0] X[c,0.9] X[c,0.9] X[c,1.2] X[j,1.3] X[c,0.8]},
  rulesep=0pt,
  stretch=0.95,
  hline{1,Z} = {1pt},
}
\midrule
QINN~\cite{Hattori2025} &
  DNN &
  M/G/1 queue model &
  Traffic-matrix scenarios &
  Flow-level &
  UDP and TCP (non-specific) &
  Avg. packet delay, throughput &
  Simulated data \\ \midrule
\cite{YuGu2004OnSimulation} &
  ns-2~\cite{1995TheNs-2} &
  Flow queueing model~\cite{Liu2003FluidNetworks} &
  Full network scenarios &
  Packet-level &
  UDP and TCP $^{\mathrm{2}}$ &
  Full packet path information &
  Simulated data \\ \midrule
MimicNet \cite{Zhang2021MimicNet:Learning} &
  ns-3~\cite{Riley2010TheSimulator} &
  LSTM~\cite{10.1162/neco.1997.9.8.1735} &
  Full network scenarios &
  Packet-level &
  Any &
  Full packet path information &
  Simulated data \\ \midrule
m3 \cite{Li2024M3:Learning} &
  flowSim~\cite{namyar2024solving} &
  Transformer (LLama2~\cite{facebookllama}) &
  Full network scenarios &
  Flow-level &
  UDP and TCP (DCTCP, TIMELY, DCQCN, HPCC) &
  99th percentile FCT &
  Simulated + Real data \\ \midrule
CausalSim \cite{Alomar2023CausalSim:Simulation} &
  Trace simulation &
  Causal DNN &
  Full network scenarios &
  Packet-level &
  Any &
  Full packet path information &
  Simulated data \\ \midrule
Sim2HW \cite{Spath2024Sim2HW:Measurements} &
  OMNET++ \cite{Varga2010OMNeT++} &
  GraphSage \cite{10.5555/3294771.3294869} &
  Full network scenarios &
  Flow-level &
  UDP &
  Min, 25th, 50th, 75th, 90th, 99th, 99.9th, 99.99th, 99.999th percentile packet delays &
  Testbed data \\
\end{tblr}
\begin{tabular}{m{\columnwidth}}
    $^{\mathrm{2}}$ Secondary model accounts only for TCP. \\
\end{tabular}
\caption{Summary of hybrid network performance models}
\label{tab:hybrid}
\end{table*}


\end{landscape}

The authors of Pantheon~\cite{Yan2018Pantheon:Research} identified this issue and proposed a model to fine-tune emulation software to track traffic performance accurately. It consists in fine-tuning emulation parameters, like propagation delay, using Bayesian optimization. Specifically, this is a process where the parameters were adjusted, evaluated using captured internet traces, and updated according to their error. This was done using the Mahimahi~\cite{191577} emulation software. Their evaluation showed a tenfold decrease in prediction error. However, this approach remains limited, due to the small number of adjustable parameters, resulting in the error rates averaging at $17\%$ for the tested traces, and the cost of emulating the different scenarios repeatedly.

A similar approach is that of iBox~\cite{Ashok2022Data-DrivenIBox}. iBox models the network as a single bottleneck link with a FIFO, drop-tail queue. iBox considers two models, discriminating between ``reactive" and ``non-reactive" cross traffic. They define non-reactive traffic as those flows whose performance is not influenced by changing the specific congestion control protocol, reactive otherwise. While the non-reactive cross traffic can be modeled directly from parameters extracted from the network traces, the reactive cross traffic parameters are learned through Bayesian Optimization and real traffic traces, as in Pantheon. The queue model is executed within a network simulator or emulator to be evaluated; in their evaluation, the authors specifically used ns-2.
iBox ultimately shares similar benefits and limitations to its predecessor: by being built with real traffic traces, it attempts to minimize the impact of training with simulated traces~\cite{Paxson1997WhyInternet, Floyd2003InternetModels, Floyd2001DifficultiesInternet}. However, the combination of Bayesian optimization’s high computational cost and the model’s simplicity may limit its practicality in real-world deployments.

\subsection{ML + Analytical Hybrid Models}

Generally, both ML and analytical models offer similar strengths and weaknesses —i.e., quick inference times but less accurate than DES. While it may be counterintuitive to combine the two, successful applications can still be found in the SotA.


An early example is Prophet~\cite{Zhang2020Prophet:Networks}, where they used ML to solve the NUM model~\cite{Kelly1998RateStability}. In the original NUM paper, the authors aimed to maximize TCP throughput, but their initial ideal model required prior knowledge of each flow's utility. In Prophet, however, they can approximate the expected utility using Srikant's unifying model~\cite{Srikant2004}. Although this too relies on another unknown, the scaling factor, Prophet approximates it through sampling and gradient descent. Sampling is done efficiently, approximating the flow parameters, and grouping flows to reduce the number of them to consider. Conversely, these modifications limit the model's applicability. For example, the flow grouping is designed assuming a Clos topology.

Another approach is to use ML to improve the reasoning of an analytical model. In~\cite{Geyer2019DeepTMA:Networks}, a GNN model is used to predict the best tandem decompositions to be used by their NC model. By acting as a heuristic, the GNN can improve the results of the NC model without incurring a significant penalty in inference time. Furthermore, the GNN model does not need to be perfectly accurate for the NC model to benefit from it. The authors expanded DeepTMA in~\cite{Geyer2020OnCalculus}, to allow the generation of decompositions, as well as performing feature analysis to understand which are the most significant features of the GNN model. While this approach reliably increases the NC model's accuracy, ultimately, it cannot address the inherent limitations present in NC models (e.g., assuming feedforward networks).

In contrast, in \cite{Helm2023PredictingGNNs}, the authors invert the dynamic, instead using the upper bounds gathered from an NC model as an additional input to a GNN model tasked to predict the performance metrics. Their analysis leveraged a GraphSage~\cite{10.5555/3294771.3294869} model, a generic GNN architecture, but its findings may be generalized to more specialized GNN architectures. They have also confirmed, as expected, that tighter NC bounds increase the accuracy of the resulting GNN model.

A similar approach was followed by the authors of QT-RouteNet~\cite{deAquinoAfonso2022QT-Routenet:Theory}. In this case, the authors use a QT model of the network —a M/M/1/B model— to extract flow and link features, which later are used as input by an RouteNet~\cite{Rusek2019UnveilingSDN} model. Doing so allowed it to generalize to topologies much larger than training: it was trained on topologies up to 50 nodes large, and evaluated in topologies ranging from 51-300 nodes. Similarly, GNNetSlice~\cite{Farreras2025GNNetSlice:Networks} also introduces network slicing information and approximate QT predictions, such as the expected maximum queuing delay and packet loss rate, as inputs to its model. The model itself is a Relational GCN (RGCN)~\cite{schlichtkrull2018modeling} designed to measure the impact of network slicing on performance. In its evaluation, it outperformed RouteNet-Fermi~\cite{Ferriol-Galmes2023RouteNet-Fermi:Networks} when predicting the average packet loss rate, delay, and jitter. 

Later, in Queue-Informed Neural Network (QINN)~\cite{Hattori2025}, integration is expanded to include a queue model in the model's loss function. In summary, a DNN model predicts both the average queuing delay and the throughput; the latter is then used by a M/G/1 queue model to obtain a second queuing delay prediction. The loss function computes the loss over both predictions. While the authors have used a DNN, this approach is compatible with other NN architectures. Ultimately, all of these approaches are a reliable way of improving the model's accuracy, but they cannot resolve ML's inherent limitations.

\subsection{Accelerated DES}

An alternative hybrid approach for network modeling, and arguably one of the most popular nowadays in the SotA, consists of reducing the computational cost of DES by replacing some elements with a faster alternative. The main objective is to maintain DES's benefits (mainly its accuracy) while minimizing its computational cost.

The first model to attempt this was in~\cite{YuGu2004OnSimulation}, where the authors combined a DES simulator (ns-2) with the fluid model in~\cite{Liu2003FluidNetworks}. The fluid model only covered TCP flows in the network's core. The paper introduces rules on how packets are updated when they cross the core, according to the values of the resolved fluid model, along with synchronization rules to avoid the DES and fluid model's states from diverging. Limitations include the fluid model's increased error rate and only applying to TCP flows. Furthermore, the translation between fluid and simulated traffic can result in predictions that generally hold but do not offer sufficient granularity. For example, the fluid model may predict accurately that a percentage of packets will drop at a given time, but cannot exactly predict which packets are dropped; instead, these will be selected at random.



It will take nearly two decades for a new iteration of this idea to be proposed, which eventually does so in the form of MimicNet~\cite{Zhang2021MimicNet:Learning}. Rather than an analytical model, MimicNet uses an LSTM model to replace parts of the network topology in the DES. The approach exploits the symmetry present in fat tree topologies, commonly used within data centers: first, it simulates a two-branch fat tree topology to train its LSTM model. During inference, it only simulates a single branch, while the rest are replaced with LSTM replicas. The LSTM models predict whether packets in the replicated branches are dropped or forwarded according to their expected behavior. By leveraging the symmetry of the topology, MimicNet remains accurate and computationally efficient. However, this results in some rigid assumptions. First, the network is assumed to be a failure-free fat tree topology, with congestion only present in the fan-in towards the flow's destination. Second, traffic patterns are expected ``scale proportionally to the size of the network", as otherwise faithful models cannot be trained using the smaller, two-branch topology.


Inspired by Parsimon~\cite{Zhao2022ScalableNetworks}, the authors of m3~\cite{Li2024M3:Learning} propose splitting the network simulation into path-level simulations. Specifically, paths are simulated separately, considering in each of them the set of foreground and background flows. Foreground flows are simulated in parallel and quickly using flowSim~\cite{namyar2024solving}, while the impact of background flows is approximated using a LLama2 transformer model~\cite{facebookllama}. These results, as well as additional scenario context (e.g., congestion algorithm), are concatenated and fed to a DNN to obtain the corrected approximations of the flow's FCT. 
The main advantage of m3 relative to Parsimon is that assuming path-level is a weaker assumption than link-level independence. Relative to its inspiration, m3's evaluation showed it to be more accurate and quicker inference times, despite including the relatively large LLama2 model. 
However, unlike most simulators, it does not maintain packet-level visibility. Instead, like Parsimon, m3 is designed for aggregated tail-prediction performance metrics, such as worst-case TCP throughput. Finally, its support of congestion control is based on being parametrized and added to the input features of its final DNN in its architecture. Consequently, it only supports those protocols seen during training.

\subsection{Simulation with DL-Enhanced Accuracy}

As we discussed, while network simulation is perceived as the most accurate approach for performance modeling, it is still subject to some inaccuracies \cite{Guemes-Palau2025RouteNet-Gauss:Learning, Yan2018Pantheon:Research}. Consequently, we have seen some approaches that attempt to use DL to enhance the simulation's accuracy to better match reality. Unlike the models in Section~\ref{sec:hybrid_models:emulation}, these are applied to simulators, not emulators, and they are complemented with DL models.

One example of this is CasualSim~\cite{Alomar2023CausalSim:Simulation}. In it, they use trace simulation, a faster yet more inaccurate alternative to DES, and instead use ML to improve its accuracy. Trace-simulation is a variant of DES where only a subset of the system is simulated, while the other segments are replaced with traffic traces. However, it assumes that the traces' contents are independent of the rest of the simulation, which rarely holds. Consequently, CasualSim proposes the use of a causal DNN model to adapt the traces according to the system behavior, improving accuracy. However, trace simulation is meant to simulate the impact of specific small changes for ``what-if" scenarios. Larger changes result in fewer elements being replaced by traces, hence becoming increasingly similar to standard DES.

Recently, in Sim2HW~\cite{Spath2024Sim2HW:Measurements}, the authors use an expanded GraphSage~\cite{10.5555/3294771.3294869} model to correct the network performance predictions given by OMNET++~\cite{Varga2010OMNeT++}. To train the model, simulated network scenarios were replicated in a testbed to obtain their ground truth.  While this approach may enhance the simulator's accuracy in replicating real-world traffic, it does not address DES's main issue: its high computational cost.

\subsection{Summary}

Ultimately, hybrid models are characterized by their diversity and practical nature. By combining existing, proven approaches, the authors of these approaches obtain stronger network models. This includes expanding emulation with network models to support performance modeling (e.g., Pantheon~\cite{Yan2018Pantheon:Research} and iBox~\cite{Ashok2022Data-DrivenIBox}), applying ML to accelerate DES (e.g., MimicNet~\cite{Zhang2021MimicNet:Learning} m3~\cite{Li2024M3:Learning}) or correct its outputs (e.g., Sim2HW~\cite{Spath2024Sim2HW:Measurements}), using ML-based heuristics to improve NC models (e.g., DeepTMA~\cite{Geyer2019DeepTMA:Networks, Geyer2020OnCalculus}), and conversely using NC and QT to improve ML model training (e.g., QINN~\cite{Hattori2025}) or to provide additional input information (e.g., \cite{Helm2023PredictingGNNs}, QT-RouteNet~\cite{deAquinoAfonso2022QT-Routenet:Theory}, GNNetSlice~\cite{Farreras2025GNNetSlice:Networks}).

The biggest benefits of these approaches are the ability to combine the strengths of both approaches. For example, when using ML to accelerate DES, ideally, the resulting model can retain DES's accuracy and granularity while benefiting from lower computational costs. However, this comes at the risk of inheriting the weaknesses as well —for example, any hybrid approach with an ML-model will require such to be trained.

\section{Discussion on Identified Trends and Challenges within Network Performance Modeling}
\label{sec:discussion}
In this section, we discuss the trends and challenges identified in current network performance models. This section is also meant to expand on earlier discussions \cite{Paxson1997WhyInternet, Floyd2003InternetModels, Floyd2001DifficultiesInternet}. 

\subsection{Balance Between Accuracy, Resolution, Applicability, and Inference Cost}
\label{sec:discussion-balance}
\begin{table*}[!t]
\centering
\resizebox{\textwidth}{!}{%
\begin{tabular}{lm{2cm}lm{3cm}m{4cm}l}
\hline
\multicolumn{2}{l}{\textbf{Approach}}           & \textbf{Accuracy} & \textbf{Expressiveness}                          & \textbf{Applicability}               & \textbf{Computational Cost}    \\ \hline
\multirow{2}{*}{Simulation} & DES      & \multirow{2}{*}{High}     & \multirow{2}{*}{Packet-level}                            & \multirow{2}{*}{General}                     & High (and sequential) \\
 & PDES &  &  &  & High  \\ \hline
\multirow{3}{*}{QT}         & Discrete & \multirow{3}{*}{Low}      & Flow-level                              & \multirow{3}{*}{Specific TCP version}        & \multirow{3}{*}{Low}                   \\
 & \multirow{2}{*}{Fluid}  & & Flow-level & &                    \\
  & & & (temporal) & &                    \\\hline
\multirow{2}{*}{NC}         & ADNC     & Medium   & Flow-level            & \multirow{2}{*}{Feed-foward networks}        & Medium                \\
 & ODNC & High & (temporal) &  & High \\ \hline
\multirow{3}{*}{ML} & \multirow{2}{*}{Shallow ML} & \multirow{2}{*}{Medium} & Flow-level & On trained topologies & \multirow{2}{*}{Low} \\
 & &  & (temporal and & and traffic profiles & \\
& DL & High  & non-temporal) & On trained traffic profiles & Medium                \\ \hline
\multirow{3}{*}{\parbox{1cm}{Hybrid \\ approaches}} &
  ML+Analytical &
  High &
  Flow-level &
  On trained traffic profiles &
  Medium \\
 &
  Accelerated DES &
  High &
  Flow-level / Packet-level &
  On trained traffic profiles; topology support varies. &
  High \\ \hline
\end{tabular}%
}
\caption{Summary of current network performance modeling approaches. }
\label{tab:qualitative_summary}
\end{table*}

An ideal network performance model should be \textbf{accurate}, \textbf{expressive} (i.e., granular predictions, ideally packet-level), \textbf{applicable} in general scenarios, and with a \textbf{low computational cost}. In practice, however, current network performance models cannot guarantee all of these properties.

This is reflected in Table~\ref{tab:qualitative_summary}, where the different approaches are summarized and qualitatively compared. 
On the one hand, DES tends to be the preferred option for network modeling, but its cost makes it unfeasible in many scenarios. PDES addresses this, not by reducing its cost, but by allowing for the simulation to be spread across more cores. This does allow it to simulate larger scenarios, but requires higher amounts of computing power, which remains impractical. As a result, research in DES and PDES focuses on reducing the cost while maintaining its other benefits.
On the other hand, both analytical and ML models tend to be computationally inexpensive. Instead, research on these models focuses on improving their accuracy, expressiveness, and the scenarios in which they are applicable while retaining the low computational cost.

Hybrid approaches, on the other hand, try to split the difference by synthesizing different approaches. For instance, accelerated DES methods seek to reduce computational cost while preserving the strengths of traditional DES. DL techniques have also proven flexible, as surveyed examples include quicker models with reasonable accuracy (e.g., RouteNet-Fermi~\cite{Ferriol-Galmes2023RouteNet-Fermi:Networks}) and more complex, complete models akin to DES (e.g., DeepQueueNet~\cite{Yang2022DeepQueueNet:Visibility}). However, current approaches still fall short of reaching all four characteristics. For example, MimicNet~\cite{Zhang2021MimicNet:Learning} is only applicable in fat tree topologies, while m3~\cite{Li2024M3:Learning} loses packet-level visibility and can only predict high percentile FCT. Furthermore, approaches such as ML+Analytical Hybrid Models tend to offer small yet significant improvements in their accuracy but do not address other fundamental constraints present in either approach.

As a result, the optimal approach will depend on its expected application. If time and computational resources are not a constraint, DES remains the best option. However, if the network performance prediction is expected to be integrated in a more complex, time-critical application (or simply a quicker, approximate prediction is desired), an analytical or ML model is better suited. Alternatively, a model that is meant to be applied only to a specific network with a static topology may not need the same degree of applicability that a general-purpose approach like DES. Ultimately, as also argued in \cite{Floyd2003InternetModels}, researchers must consider their use case when deciding which measurements to use as input, which aspects of the network should be modeled in this use case, and which properties are most relevant.
The differences in approaches are also reflected in how the field tries to develop better network models. While some researchers focus on reducing the costs of the heavier, more accurate models, others try to improve the accuracy of the more efficient ones.

Furthermore, it is not unreasonable to believe that such an ``ideal" model is impossible to build to begin with, as the different features can impede others. For example:
\begin{itemize}
    \item An expressive model should predict the behavior of individual packets, yet their amount in modern networks is measured in the billions~\cite{Guemes-Palau2025RouteNet-Gauss:Learning}. Hence, obtaining packet predictions, without aggregation or summarization, will be inherently expensive by sheer scale.
    \item Similarly, aggregation also implies losing information as the sequences of packets get shortened to a fixed set of values, resulting in a potential loss of accuracy.
    \item Analytical and statistical models rely on assumptions for them to be accurate. However, these may constrain the range of scenarios they can be applied to. Nonetheless, without such assumptions, the models fail to characterize network traffic and its behavior (no-free-lunch theorem~\cite{ho2002simple}).
\end{itemize}

\subsection{The Dominance of DES and the Surge of GNNs}
\label{sec:DES_GNN_Dominance}

Traditionally, DES has been regarded by network operators as the gold standard for network performance modeling. This is exemplified by the success of major DES simulators like ns and OMNET++, as well as the fact that most non-DES network performance models are compared using or against simulated data. Referring back to Figure~\ref{fig:evolution} in the introduction, simulation is the only type of model that has had constant attention over the last three decades. This can be explained by the fact that simulation is both one of the most accurate options and capable of offering packet-level predictions.

However, since 2018, we have seen a surge in DL network performance models. This is in part due to the success of the GNN architectures. Their design exploits the relational information in computer networks to their advantage, proving to be accurate and computationally inexpensive. While they do not offer packet-level granularity, they can be augmented to include a temporal resolution~\cite{Guemes-Palau2025RouteNet-Gauss:Learning, Huang2024XNet:Networks, Li2025M4:Simulator}. This has resulted in their dominance as the most used DL architecture when building network models: 17 out of the 29 ($\approx58\%$) pure ML models identified in the survey are based on GNNs. They are also quite relevant in hybrid approaches: 5 out of the 12 hybrid models identified included a GNN model.

Note that this prevalence is not universally shared across the entire community. Let us consider models published since 2018 in ACM SIGCOMM and IEEE INFOCOM, the two CORE A* conferences regarding computer networking. 
On the one hand, the IEEE INFOCOM does reflect the prevalence of GNNs, with two out of three models accounted for being GNN-based~\cite{Ferriol-Galmes2023RouteNet-Fermi:Networks, Huang2024XNet:Networks} and the remainder one being a hybrid method involving ML~\cite{Zhang2020Prophet:Networks}. On the other hand, ACM SIGCOMM instead prefers PDES and similar methods: out of the seven models published since 2018, three models are PDES proposals~\cite{Gao2023DONS:Parallelization, Li2022SimBricks:Simulation,Khashab2025NSX:Server}, and another two models are DL-accelerated DES hybrid approaches~\cite{Zhang2021MimicNet:Learning, Li2024M3:Learning}. Of the remaining two models, only one is a GNN~\cite{Suarez-Varela2019ChallengingModeling}. The other, DeepQueueNet~\cite{Yang2022DeepQueueNet:Visibility}, is a transformer-based model whose design mimics DES reasoning, down to reasoning over individual packets. These discrepancies show that different voices in the communities may show preferences in which aspects of network performance simulation they prioritize, hence preferring approaches whose strengths align with them.

\subsection{Reduced Interest in Analytical Models}
\label{sec:analytical_end}

While there has been a rise in DL-based models, specifically the GNN architectures, it has come at the cost of slowing the development of newer analytical models. Out of the models we have surveyed, only one purely analytical model was published in the last 5 years~\cite{Fiorini2025}, with the previous one being published in 2017~\cite{Bondorf2017QualityAnalysis}. We have identified some potential reasons why.

First, the most straightforward explanation is the fact that ML (and DL) models offer similar benefits while outperforming analytical models. At first, the main difference that analytical models offered against DES was their reduced computational cost, at the cost of less expressive and accurate results. Nowadays, ML models also offer accurate and computationally inexpensive predictions, hence placing themselves in direct comparison against analytical models.


Second, unlike ML, analytical models try to explicitly define complex network behavior through sets or systems of equations.
ML models either treat networks as black boxes and try to predict their performance through regression (e.g., \cite{He2005OnThroughput} and CLAAP~\cite{Monaco2025Real-timeApplications}), or explicitly represent some existing dependencies within the network but still rely on the training process for them to be completely developed (e.g., RouteNet~\cite{Rusek2020RouteNet:SDN}). By contrast, analytical models must be completely formulated by their creators, which involves defining their assumptions and mathematically proving their validity or error bounds. In turn, this makes analytical models more rigid, less future-proof, and arguably harder to build overall.

An example of this is how different approaches support multiple versions of TCP simultaneously. This is an important aspect for network models, as TCP implementations evolve, and it is not uncommon for several versions to co-exist in the same network \cite{Paxson1997WhyInternet}. In analytical models, authors either assume a ``generic" TCP version, which results in inaccuracies as it fails to capture the differences between implementations \cite{Marsan2004UsingNetworks, Baccelli2002AIMDTraffic, Bohacek2003ATransmission}, or are forced to re-formulate segments of the model for each version they support \cite{Bonald1999ComparisonFairness, Liu2003FluidNetworks}. In contrast, DL models that support multiple TCP versions usually differentiate between versions through a one-hot encoded vector \cite{Jaeger2022ModelingNetworks, Li2025M4:Simulator}. This means that DL models can be expanded to fit more TCP versions easily, as long as the authors have the recorded scenarios to train them with. 

There is also the fact that, as new DL architectures keep being developed, researchers will continue to explore their potential as network performance models, as it happened recently with the transformer architecture \cite{Dietmuller2022AGeneralization, Yang2022DeepQueueNet:Visibility}. Altogether, this can explain why the research into new analytical models has slowed down, and instead, this effort has moved into the development of ML models.

\subsection{Adapting to Changing Networks}
\label{sec:discussion_new_protocols}
One of the biggest challenges of network performance modeling identified early on was that of describing the internet as an ``immense moving target" \cite{Paxson1997WhyInternet, Floyd2001DifficultiesInternet}. They identified the dynamic nature of networks, how they change over time in terms of traffic patterns, usage, and implemented protocols, and how these could be a challenge to network models at the time. This remains an ongoing challenge: since the 2000s, internet usage has continued to increase, data centers rely on newer TCP variants like DCTCP~\cite{10.1145/1851182.1851192} and DCQN~\cite{10.1145/2829988.2787484}, and wireless networks are more common and complex.

At the time, solutions proposed to address this challenge were based on the modeling approaches prevalent at that time. Analytical models were seen as the most vulnerable, as they tend to rely more on network assumptions or, in the case of queuing and fluid models, they were designed with a given TCP version in mind. A solution proposed back then was searching for invariant properties —aspects or properties that remain constant over different network topologies, sizes, and usages~\cite{Paxson1997WhyInternet, Floyd2001DifficultiesInternet}. Examples include the self-correlation in packet inter-arrival distributions, or the heavy-tail distributions present in metrics such as packet delay, RTT, or FCT.

Another solution proposed was the push for modular, interoperable models~\cite{Floyd2003InternetModels}. The idea was to avoid building monolithic network models capable of individually addressing any scenario. Instead, it proposed building network models for them to be interoperable, that is, for them to be combined and fed to each other. This allows models to be gradually updated to newer developments, like newer TCP versions. This approach is better suited for DES simulators, as these can be expanded to cover new protocols and devices as they are released. Open-sourced simulators like ns~\cite{1995TheNs-2, Riley2010TheSimulator} and OMNeT~\cite{Varga2010OMNeT++} have remained updated thanks to the community. Furthermore, we have modular PDES like SimBricks~\cite{Li2022SimBricks:Simulation} and SplitSim~\cite{Li2024SplitSim:Research} whose modularity enables their coverage of network features as well as their parallelization. Finally, DL models do offer new ways to address this issue, we will cover it later in Section~\ref{sec:future_tl_adapt}.


\subsection{Simulation-Dominated Evaluation}
\label{sec:simulation_evaluation}
\begin{figure}
    \centering
    \includegraphics[width=0.8\columnwidth]{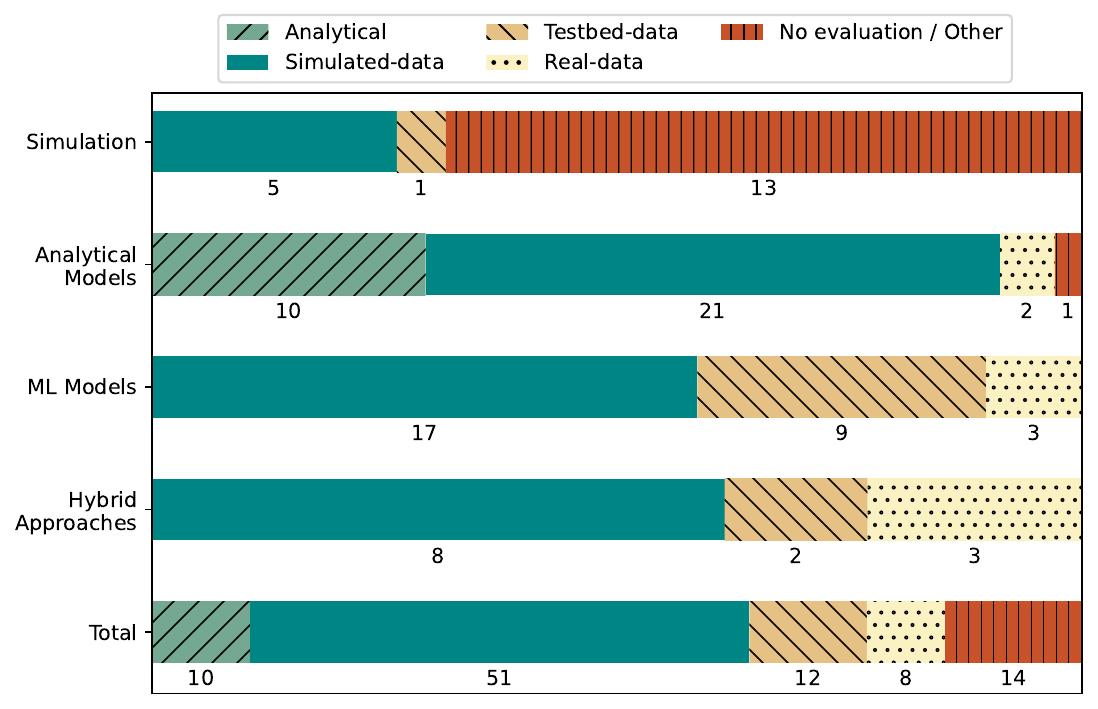}
    \caption{Evaluation categories across different model types. If a model is evaluated with multiple categories, they are categorized in the following priority: real data, testbed data, simulated data, and analytical.}
    \label{fig:evaluations}
\end{figure}

Another trend in network modeling is the prevalence of using simulated data in the evaluation. This fact is reflected in Figure~\ref{fig:evaluations}, which shows how models across the different approaches are evaluated. Overall, it shows that the use of simulated data dominates, used in over half of the surveyed models, and is the most used across all the approaches except simulation itself; even across simulators, comparing against other simulators remains the preferred evaluation choice. This is because most simulators surveyed are published through white papers, which describe the simulation implementation but otherwise do not measure their accuracy.
Otherwise, we see a minority of models being evaluated with testbed or captured data in ML and hybrid approaches, while another significant minority of samples are just evaluated analytically in the case of the analytical models. 

Overall, there are strong reasons for this. First, captured data is scarce and not always available. Unlike it, simulated data can be generated to cover any desired scenario, allowing for more diverse datasets. Second, captured data may also be limited by the features that the data's authors were able to capture, while in simulation, the entire state of the network and traffic is present. Furthermore, testbed networks require a monetary cost to set up and later upgrade (e.g., changing devices). Meanwhile, simulation software can be run on generic computing devices without additional hardware.
However, simulation has several disadvantages, as commented back in Section~\ref{sec:simulation}. First, its high computational cost limits the size of the networks or traffic intensities used in the evaluation. Hence, whenever the model is applied to these challenging scenarios, it may perform worse than expected. There is also the issue where simulation may not cover specific protocols and network devices, or without the required accuracy~\cite{Floyd2001DifficultiesInternet, 10.1145/268437.268737, Yan2018Pantheon:Research, Guemes-Palau2025RouteNet-Gauss:Learning}.

\subsection{Heterogeneous Approaches, Heterogeneous Evaluations}

Another challenge present in network performance modeling is a lack of common evaluation procedures. We believe that this is due to the heterogeneity of modeling approaches compounded by the scarcity of public datasets. The differences in network performance modeling approaches allow for the creation of models with different strengths and goals. Such versatility permits network operators to choose the modeling approach that best fits their needs. First, models may differ on the performance metric they measure, which depends on the traffic type they are modeling (e.g., traffic from a specific protocol like TCP). This also influences what data they take as input, as some use solely traffic traces, while others can include the entire network in their reasoning. Later, some models can be only applicable under certain constraints (e.g., NC only applicable to feed-forward networks, or MimicNet~\cite{Zhang2021MimicNet:Learning} only to fat tree topologies), which also limits common scenarios where they can be compared to other models. Even when considering the same metric, the granularity of their outputs may also condition how models are evaluated. For example, models that predict on a packet-level, or even a flow-level with a temporal component, must see their predictions aggregated to be compared against models offering flow-level predictions. 

Furthermore, even comparing packet-level outputs against each other may be complicated. For example, one model may predict a given delay for a packet that, according to the ground truth, was lost. This, in turn, makes us rely on error metrics like the Wasserstein metric to compare the distribution of the predictions. While useful, such metrics lack the intuitiveness behind the obtained result. That is, unlike metrics like the Mean Absolute Percentage Error that can be easily understood, there is no clear way of interpreting whether a given Wasserstein metric can be regarded as a ``good" or ``bad" result. At most, it can be used to compare models, knowing that lower values mean closer predictions to the ground truth.

Note that these difficulties also apply to models that offer flow-level predictions with a temporal component. This may be exacerbated by the fact that these models may consider their temporal component under different scales (e.g., predictions every second versus every millisecond). Even worse, models may define their temporal components differently. For example, while RouteNet-Gauss~\cite{Guemes-Palau2025RouteNet-Gauss:Learning} may consider fixed-length windows, the temporal component in m4~\cite{Li2025M4:Simulator} is event-driven.

Finally, all of these difficulties are worsened by the lack of publicly available network data. This is the case for several reasons. In the case of real-world captured data, only a few groups and companies have access to such, and the capability to capture it. Furthermore, publishing real data from real users may pose privacy risks and requires anonymization procedures for it to be safe to be made public. Alternatively, testbed networks are rare due to their costs; hence, by extension, published datasets will also be rare. 

\section{Future Directions and Opportunities}
\label{sec:future}
In this section, we follow up on the previous discussion with our prediction on future research directions for network performance modeling. This is based on our expected evolution of current trends and how current challenges may be addressed.

\subsection{Consolidation of PDES and GNNs Models}

First, from the conclusions obtained from Sections~\ref{sec:DES_GNN_Dominance} and~\ref{sec:simulation_evaluation}, we can derive that DES is one of, if not the most desirable, modeling approaches available to network operators nowadays. Unsurprisingly, lately, there has been a push by researchers to address the challenge of PDES design. Hence, we believe that PDES research will continue to consolidate its position as the new ``baseline" method, replacing traditional (single-process) DES simulators. This means reducing the synchronization overhead while retaining correctness. We believe that the increased availability of computing power~\cite{IEADatacenter} means that computational efficiency will not be as much of a priority compared with the ability of the PDES to exploit the resources available. The most recent PDES surveyed, NSX~\cite{Khashab2025NSX:Server}, exemplifies this, proposing a highly distributed simulation that can run in the same specialized hardware used to train AI models.

By analyzing the SotA, we can also conclude that DL models, and specifically GNNs, are well-positioned to become the main alternative to simulation.
They can be applied to scenarios where PDES cannot, such as those with restrictions on computing resources available. Then, for the remaining scenarios, DL models are magnitudes of times quicker than PDES while also benefiting from parallelization and specialized hardware thanks to libraries like Tensorflow and Pytorch. Among DL architectures, specialized GNNs for networking have proved to be the most effective, being both cost-effective, accurate, and more robust than their other alternatives.

However, there is still progress to be made for GNNs. For example, support for congestion control algorithms~\cite{Jaeger2022ModelingNetworks, Geyer2019DeepComNet:Learning} or the temporal component~\cite{Guemes-Palau2025RouteNet-Gauss:Learning} is still fairly recent. Also, current models can only faithfully predict traffic patterns seen during their training. Ultimately, addressing these issues will increase the number of scenarios GNNs can be applied to and establish them as a reliable alternative to DES.

\subsection{Analytical Models Enhancing DL Models}

In Section~\ref{sec:analytical_end}, we discussed how research into analytical models is getting reduced attention, as ML and DL network models are proving to be more cost-effective. However, we have seen how analytical models are still being used in conjunction with other techniques. An example of this is how both QT and NC models have been used as heuristic input to be fed to a DL model to improve their accuracy~\cite{Helm2023PredictingGNNs, deAquinoAfonso2022QT-Routenet:Theory}. Recently, studies have also shown that they can be used to define more informed loss functions, further improving the training process~\cite{Hattori2025}.

Ultimately, analytical models still offer inexpensive, good approximations of the expected performance that more complex network models can later refine. The proposed approaches also benefit from the fact that they can be applied independently of the underlying ML architecture, easing their implementation. Further research in this area can lead to improved ML models that retain their cost-effectiveness. This includes studying other aspects in which the analytical model can be integrated into the DL architecture beyond the input features and loss functions, as well as how to incorporate more complex analytical models without incurring excessive computational costs.

\subsection{ML as a New Tool for Evolving Networks}
\label{sec:future_tl_adapt}

Section~\ref{sec:discussion_new_protocols} discussed the challenge of the ever-changing nature of networks, and how to address models becoming outdated because of it. Proposals in the past include models focusing on true invariant properties of networks or designing systems of interoperable models. 

In addition to these proposals, we believe ML, and specifically DL, has unlocked a new way to handle changing networks.
Unlike analytical models or simulations, an ML model does not explicitly encode the network dynamics. Instead, it learns these dynamics from its training data. Hence, under the assumption of having an appropriate ML architecture to learn such network dynamics, such as a GNN, it is reasonable to believe that the architecture itself does not need to be modified to be adapted to changes in the network. Instead, the ML model would have to be re-trained, a process that, while requiring some computational effort it no longer requires the expert knowledge required to adapt an analytical model, implement a simulator, or design an ML architecture.

We refer to this new approach as \textit{design once, train as necessary}, and it has already been proposed in papers such as~\cite{Dietmuller2022AGeneralization}. Besides adapting to evolving network conditions, it would also allow for designing a general architecture valid for many specific use cases, depending on how it is trained. Furthermore, re-training may be less costly than training from scratch by exploiting transfer learning, as done in existing works~\cite{Li2024GLANCE:Networks, Hattori2024MetaMetrics, Hattori2025MetaMetrics}. At its best, this approach allows the advantages of using a universal design for building network performance models, while making their implementation specific for each use case, adapting to its necessities.

Nonetheless, this approach still has some drawbacks that future research must address. First, with the current formulation, the model would still require new traffic measurements for its readjustment. While transfer learning would reduce the amount of data needed, it can still be a costly process. Also, until the model is adjusted, it cannot be expected to work accurately. Hence, DL models should still be designed to be as generalizable as possible, to reduce the amount of retraining to be done. 

Another risk is the gradual degradation of the model's accuracy introduced by minor changes, rather than a single significant change. The issue with this is that the gradual degradation may be harder to identify, and hence, the operators may be working with an inaccurate model without realizing it. This may be adjusted with a continuous model evaluation, but requires periodic network measurements.

\subsection{Data Center-Centric Designs}

The rise in data center demand, and its focus on ML-related workloads, is an opportunity to develop more specific, data center-centric network models. Specifically, data centers already share common, well-researched topologies (e.g., Fat Trees~\cite{6312192}), and currently it is expected that future demand will be mainly due to the training and usage of Large Language Models or similarly large DL models~\cite{IEADatacenter}.
Altogether, most wired network scenarios in the future will likely represent networks with similar topologies, hardware, and even traffic patterns as they will be dedicated to the same ML-related use cases. 

Consequently, this homogenization introduces common properties across data center networks. These can be leveraged, in the same way as other invariant properties, to simplify model design without compromising model accuracy. While such models are highly specialized, given the rising demand and importance of data centers, it is a sensible trade-off.

Among the surveyed models, we have already found models that do benefit from the increased demand in data centers. MimicNet~\cite{Zhang2021MimicNet:Learning} is a powerful network model, being accurate, cost-effective, and granular, but only applicable to Fat Tree topologies. NSX~\cite{Khashab2025NSX:Server} is a network simulator that is purposefully built to take advantage of modern data centers — that is, being designed to run concurrently in multiple GPUs.

\subsection{Better Usage of Simulation Data for Training Real-World Models}

Back in Section~\ref{sec:simulation_evaluation}, we discussed the prevalence of DES in the evaluation of network models and the issues arising from doing so. It is worth noting that evaluating models (and training them in case of ML-based ones) on simulated data may not lead to accurate results when addressing real-world traffic. Hence, future work must find ways to reduce this discrepancy.
Currently, one way this may be addressed is by improving on DES itself. For example, more efficient PDES simulators can be used to simulate scenarios too large for traditional DES to handle. There is also work like~\cite{Spath2024Sim2HW:Measurements} where the DES's output is corrected using a surrogate ML model.

Another popular approach, specifically when building ML models, is framing the discrepancies between simulation and reality as a transfer learning problem.
Transfer learning is a series of techniques meant to exploit trained models for a given task to assist in the training of models for a second, related task. In this case, transfer learning would consist of using ML models trained with simulated data to assist models to be applied to real-world traffic. This allows for the latter to require fewer samples to build and benefit from additional accuracy. Among the surveyed works, in~\cite{Hattori2024MetaMetrics, Hattori2025MetaMetrics} simulated samples are effectively used through transfer learning, but it relies on a specific architecture that benefits from it. In the recently published~\cite{guemes2025bridging}, transfer learning was successfully applied to a RouteNet-Fermi~\cite{Ferriol-Galmes2023RouteNet-Fermi:Networks} model.


\section{Conclusions}
\label{sec:conclusions}
In conclusion, this survey analyzes the evolution of network performance modeling over the last decades. We have identified 95 unique network performance models spread across multiple conferences and journals. By identifying the taxonomy of modeling approaches, we can gain a deeper understanding of the evolution of priorities within the research and professional community. For example, we observed an evolution in preferred methodologies, as models have transitioned from analytical models to ML and hybrid approaches.


From the surveyed models, we have recognized the properties sought after by network operators: accuracy, expressiveness (the level of detail in the results), applicability in any plausible scenario, and low computational cost. While different approaches focus on one or several of these properties, no surveyed network performance model can achieve all of them simultaneously. Admittedly, it may be impossible for all of these properties to be reached simultaneously, as they impede each other. For example, accurate and expressive models tend to be more complex, which will be more costly.

Consequently, this leads to the heterogeneity and diversity of the available approaches in network performance modeling's SotA. On the one hand, such heterogeneity allows the design of specialized models, allowing researchers to optimize those properties that may be most relevant to the scenario at hand. On the other hand, it also makes the comparison between approaches harder, which is then compounded by the limited availability of public networking datasets. 
We have also discussed other open problems in network performance modeling. First, we have the issue of ever-evolving networks, a problem posed over 20 years ago, which still challenges the ability to build models applicable to future networks. Second, the majority of the identified network performance models rely on simulated data for their evaluation, which may compromise their expected effectiveness when applied in real-world scenarios. 

Finally, we have identified potentially fruitful research directions that are just starting to be explored. ML-based models and transfer learning are a promising approach to address changing networks. Advances in PDES may result in feasible simulations of large network topologies. Furthermore, we expect that other trends, like the increased demand for data centers, will shape future research.


\section*{Acknowledgments}
This publication is part of the I+D+i project titled BLOSSOMS, grant PID2024-158530OB-I00, funded by MICIU/AEI/10.13039/501100011033/ and by ERDF/EU.
This work is also partially funded by the Catalan Institution for Research and Advanced Studies (ICREA). 
Carlos Güemes is funded by the AGAUR-FI ajuts (Grant Ref. 2023 F-1 00083) Joan Oró of the Secretariat of Universities and Research of the Department of Research and Universities of the Generalitat of Catalonia and the European Social Plus Fund.

\section*{CRediT authorship contribution statement}

\textbf{Carlos Güemes-Palau:} Conceptualization, Investigation, Visualization, Writing - Original Draft, Writing - Review and Editing \textbf{Miquel Ferriol-Galmés:} Writing - Original Draft, Writing - Review and Editing \textbf{Jordi Paillisse-Vilanova:} Writing - Original Draft, Writing - Review and Editing \textbf{Pere Barlet-Ros:} Writing - Review and Editing, Supervision \textbf{Albert Cabellos-Aparicio:} Supervision, Writing - Review and Editing, Funding acquisition

\section*{Declaration of competing interest}
Carlos Güemes-Palau reports financial support was provided by Spain Ministry of Science and Innovation. Miquel Ferriol-Galmés reports financial support was provided by Spain Ministry of Science and Innovation. Jordi Paillisse-Vilanova reports financial support was provided by Spain Ministry of Science and Innovation. Pere Barlet-Ros reports financial support was provided by Spain Ministry of Science and Innovation. Albert Cabellos-Aparicio reports financial support was provided by Spain Ministry of Science and Innovation. Carlos Güemes-Palau reports financial support was provided by Generalitat de Catalunya Ministry of Research and Universities. Carlos Güemes-Palau reports financial support was provided by European Social Plus Fund.  Pere Barlet-Ros reports financial support was provided by Catalan Institution for Research and Advanced Studies. Albert Cabellos-Aparicio reports financial support was provided by Catalan Institution for Research and Advanced Studies. If there are other authors, they declare that they have no known competing financial interests or personal relationships that could have appeared to influence the work reported in this paper.

\section*{Data availability}
No data was used for the research described in the article.


%

\bibliographystyle{elsarticle-num}
\bibliography{mendeley_references,additional_references}

\end{document}